\pdfoutput=1
\documentclass[11pt,letterpaper]{article}
\usepackage[T1]{fontenc}
\usepackage[utf8]{inputenc}
\usepackage[british]{babel}
\usepackage{lmodern}
\usepackage[letterpaper,margin=1in]{geometry}
\usepackage{amsmath,amssymb,amsthm,mathtools}
\usepackage{aliascnt}
\usepackage{microtype}
\usepackage{enumitem}
\usepackage{booktabs,array}
\usepackage{flafter,placeins,needspace}
\usepackage{xcolor}
\usepackage{tikz}
\usetikzlibrary{arrows.meta,positioning,calc,patterns,fit,decorations.pathreplacing}
\definecolor{aliceblue}{HTML}{2673A6}
\definecolor{boborange}{HTML}{D77A24}
\definecolor{carolgreen}{HTML}{38956E}
\definecolor{residuegray}{HTML}{7F8994}
\usepackage[unicode,colorlinks=true,linkcolor=blue!45!black,citecolor=blue!45!black,urlcolor=blue!45!black]{hyperref}
\usepackage[capitalise,nameinlink,noabbrev]{cleveref}
\numberwithin{equation}{section}
\newtheorem{theorem}{Theorem}[section]
\newaliascnt{lemma}{theorem}
\newtheorem{lemma}[lemma]{Lemma}
\aliascntresetthe{lemma}
\newaliascnt{proposition}{theorem}
\newtheorem{proposition}[proposition]{Proposition}
\aliascntresetthe{proposition}
\newaliascnt{corollary}{theorem}

\aliascntresetthe{corollary}
\newtheorem{maintheorem}{Theorem}
\newaliascnt{mainproposition}{maintheorem}
\newtheorem{mainproposition}[mainproposition]{Proposition}
\aliascntresetthe{mainproposition}
\theoremstyle{definition}
\newaliascnt{definition}{theorem}
\newtheorem{definition}[definition]{Definition}
\aliascntresetthe{definition}
\theoremstyle{remark}
\newaliascnt{remark}{theorem}
\newtheorem{remark}[remark]{Remark}
\aliascntresetthe{remark}
\newtheorem{introremark}{Remark}[section]
\crefname{maintheorem}{Theorem}{Theorems}
\Crefname{maintheorem}{Theorem}{Theorems}
\crefname{mainproposition}{Proposition}{Propositions}
\Crefname{mainproposition}{Proposition}{Propositions}
\crefname{introremark}{Remark}{Remarks}
\Crefname{introremark}{Remark}{Remarks}
\newcommand{\E}{\mathbb{E}}

\DeclareMathOperator{\osc}{osc}

\DeclareMathOperator{\Cut}{Cut}
\DeclareMathOperator{\Eval}{Eval}

\setlist{topsep=4pt,itemsep=2pt,parsep=0pt}
\hypersetup{pdftitle={Exact Envy-Free Cake Cutting with Polynomial Expected Queries},pdflang={en-GB},pdfsubject={Envy-free cake cutting; randomised oracle computation},pdfauthor={Simon Mackenzie and Mashbat Suzuki}}

\renewenvironment{abstract}{\begin{center}\bfseries Abstract\end{center}\noindent\ignorespaces}{\par}
\title{Exact Envy-Free Cake Cutting with\\Polynomial Expected Queries}
\author{Simon Mackenzie\\UNSW Sydney\\\texttt{simon.william.mackenzie@gmail.com}
\and Mashbat Suzuki\\UNSW Sydney\\\texttt{mashbat.suzuki@unsw.edu.au}}
\date{October 2026}
\begin{document}
\hypersetup{pageanchor=false}
\begin{titlepage}
\maketitle
\begin{abstract}
Envy-free cake cutting asks how to divide a resource so that no agent prefers another agent’s share. Finding such a division with few queries about preferences has been a longstanding challenge. Aziz and Mackenzie gave the first query bound depending only on the number $n$ of agents: a tower of six $n$’s. Subsequent improvements culminated in Ye and Bai’s single-exponential bound of $n^{O(1)}2^n$. A large gap remains between this deterministic upper bound and Procaccia’s general lower bound of $\Omega(n^2)$. Whether polynomially many queries suffice is a central open question in fair division.

We show that a polynomial expected query bound is possible. Our randomised protocol uses only independent, uniformly sampled points from the cake as its source of randomness. For every profile of $n$ agents with nonnegative, additive, nonatomic valuations, the protocol has the following guarantees:
\begin{itemize}
\item It terminates with probability one.
\item Every allocation it returns is complete and exactly envy-free.
\item Its expected number of Robertson--Webb queries is $\widetilde{O}(n^{30})$.
\end{itemize}
The expectation is taken solely over the protocol’s random choices, with no distributional assumptions on preferences.
\end{abstract}

\thispagestyle{empty}
\end{titlepage}
\setcounter{page}{1}
\hypersetup{pageanchor=true}
\section{Introduction}\label{sec:intro}

Fair division asks how to share a resource among agents who may value its parts differently. A natural fairness requirement is \emph{envy-freeness}, which means that each agent values her own share at least as highly as anyone else’s. Finding such a division requires information about the agents’ preferences. The algorithmic challenge is to determine how much information is needed.

We study this question in cake cutting, where a divisible resource is represented by the interval $[0,1]$. Each agent assigns finite, nonnegative values to pieces. These valuations are additive and nonatomic: values of disjoint pieces add, and the value of an interval varies continuously with its endpoints. In particular, individual points have zero value. We assume that every agent assigns positive value to every interval of positive length; such agents are called \emph{hungry}. These interval values may be arbitrarily small and may differ greatly across agents.

The protocol learns about these valuations through two types of query: an $\Eval$ query asks an agent for the value of an interval, and a $\Cut$ query asks her to identify a cut that yields a specified value. These are the standard Robertson--Webb queries \cite{RobertsonWebb1998,WoegingerSgall2007}, and their answers are exact. Hungriness ensures that every feasible $\Cut$ query has a unique answer, simplifying the protocol’s description. Our goal is to find an exactly envy-free division of the \emph{whole} cake, allowing each share to consist of several intervals. We seek a query bound that depends only on the number of agents, regardless of their valuations.

Obtaining a bound of this kind was a major step in the study of envy-free cake cutting. Brams and Taylor~\cite{BramsTaylor1995} gave a protocol that terminates for every profile of additive, nonatomic valuations, but whose query count is not bounded by any function of the number of agents. Aziz and Mackenzie obtained bounded protocols first for four agents~\cite{AzizMackenzie2016a} and then for any number of agents~\cite{AzizMackenzie2016}. Amanatidis et al.~\cite{Amanatidis2018} later improved the four-agent bound.

For general $n$, however, Aziz and Mackenzie’s bound was a tower of six $n$’s, $n^{n^{n^{n^{n^n}}}}$. Subsequent work substantially reduced this bound: Sokolov obtained $n^{8n^2(1+o(1))}$~\cite{Sokolov2023}, and Ye and Bai recently improved it to $n^{O(1)}2^n$~\cite{YeBai2026}. Even so, a large gap remains between the best known deterministic bounds: the upper bound is single-exponential, whereas Procaccia’s lower bound is $\Omega(n^2)$~\cite{Procaccia2009}. A central open question in this line of work is whether polynomially many queries suffice to find an exactly envy-free division.

\bigskip

\noindent\textbf{Our Contribution.} We make progress on this question by giving a randomized protocol with access to an oracle that returns independent uniform samples from $(0,1)$. 
Since the cake is represented by $[0,1]$, each oracle call simply selects a point uniformly at random from the cake. With this source of randomness, the protocol has the following guarantees:
\begin{itemize}
\item It terminates with probability one.
\item Every allocation it returns is complete and exactly envy-free.
\item Its expected number of Robertson--Webb queries is polynomial in the number of agents.
\end{itemize}
The same polynomial bound holds for every valuation profile. The expectation is over the protocol's random choices, with the valuations fixed; we assume no distribution over preferences. The following theorem makes these guarantees precise.

\Needspace{9\baselineskip}
\begin{maintheorem}[Complete exact envy-free division]\label{thm:main}
There exists a randomised protocol that, for every $n\ge1$ and every profile of $n$ hungry agents with nonnegative, additive, nonatomic valuations,  returns a complete, exactly envy-free allocation using $\widetilde{O}(n^{30})$ Robertson--Webb oracle calls in expectation. The expectation is over the protocols random choices.
\end{maintheorem}

The expected number of independent uniform draws from $(0,1)$ is also $\widetilde{O}(n^{30})$.

We also obtain a polynomial bound on the expected total computation between oracle calls.

\begin{mainproposition}[Expected computation]\label{prop:main-computation}
The protocol in \cref{thm:main} can be implemented so that, for every
$n\ge1$ and every profile of $n$ hungry agents with nonnegative,
additive, nonatomic valuations, it returns a complete, exactly
envy-free allocation using $n^{O(1)}$ operations in expectation.
The expectation is over the protocol's random choices.
\end{mainproposition}

In  \Cref{prop:main-computation}, we count each Robertson--Webb query, independent uniform draw from $(0,1)$, exact arithmetic operation, or comparison as a single operation.

We also give a protocol whose query bound depends on the requested confidence and holds on every run. For any $\delta\in(0,1)$, it returns a complete, exactly envy-free allocation with probability at least $1-\delta$ and reports failure otherwise; it never returns an allocation that violates envy-freeness. The full guarantee is stated in \cref{prop:continuous-capped}.

\bigskip

\begin{introremark}[The hungry-agent assumption]\label{rem:general-valuations}
We assume hungriness only to simplify the presentation. Both main results, \Cref{thm:main} and \Cref{prop:main-computation}, remain valid without this assumption: valuations with zero-valued intervals can be handled using suitable conventions for zero values and cut points.
\end{introremark}

\paragraph{Relation to earlier protocols.}
The difficulty is not just to produce an envy-free partial division.
The cake left over must eventually be allocated without creating envy.
Earlier protocols give some agents enough of an advantage to protect
them against any later division of the remainder
\cite{BramsTaylor1995,AzizMackenzie2016}. Ye and Bai organise this
completion argument using polynomially many saved partial allocations.
Their analysis identifies the exponential cost of calls to Core and
SubCore, procedures introduced by Aziz and Mackenzie~\cite{AzizMackenzie2016},
as the remaining obstacle to a polynomial bound~\cite[Section~5]{YeBai2026}.
Our local procedures supply the guarantees needed for completion with
polynomial expected cost. We retain the saved allocations, useful
reassignments and eventual protection against the remainder. The key
repair procedure removes envy by reassigning whole bundles and adding a
small amount of cake. Applied to all agents still participating, it
preserves enough of a selected group's large advantages for that group
to keep its shares while the others divide the remainder.

Our new ingredient is a way to compensate agents with cake while
controlling the envy that those payments could create. It builds on
Halpern and Shah's envy-removing monetary subsidies~\cite{HalpernShah2019}
and the use of cake to realise payments in Aziz, Lu, Mackenzie and
Suzuki~\cite{AzizLuMackenzieSuzuki2025}. Their construction uses cake pieces
that all agents value in prescribed common proportions. Our payments may
have different values for different agents: we make each recipient's
payment exact and bound the discrepancies for everyone else.
This allows compensation with polynomially many queries even when agents'
preferences are arbitrarily close. Randomisation makes the necessary
accuracy affordable.

\paragraph{Scope and related guarantees.}
Ch\`eze obtains a polynomial query bound with high probability when values
on a fixed partition follow a random-input model~\cite{Cheze2026}. Our
expected bound holds with the valuations fixed; we randomise
the protocol. Mehra and Psomas give a deterministic $O(n^4)$-query protocol
for a fairness requirement that strengthens proportionality but is strictly
weaker than envy-freeness for $n\ge3$~\cite{MehraPsomas2025}.
Free-disposal protocols may leave cake
unallocated while retaining exact envy-freeness~\cite{SegalHaleviHassidimAumann2016}. Stromquist's
finite-protocol impossibility requires each share to be one interval~\cite{Stromquist2008}.

Randomised lower bounds also depend on the query and allocation model.
Arndt, Pruhs and Tran prove an expected $\Omega(n\log n)$ query lower
bound for proportional division when each agent must receive a single
interval~\cite[Sections~1.3 and~2]{ArndtPruhsTran2026}.
Edmonds and Pruhs obtain a randomised lower bound when cut replies permit
a fixed relative error~\cite[Theorem~3.4]{EdmondsPruhs2011}. Neither result directly
applies to our combination of exact cut queries and possibly disconnected
shares.

After specifying the model in \cref{sec:model}, we state the three local
guarantees and illustrate the payment idea in \cref{sec:overview}.
\Cref{sec:completion} uses those guarantees to complete the division;
\cref{sec:kernel,sec:snapshot} construct the local procedures.
\Cref{sec:accounting} combines their costs.
\section{The query and computation model}\label{sec:model}

\subsection{Cake, values and allocations}

The cake is $[0,1]$, and a \emph{piece} is a finite union of intervals.
Agent $i$ assigns a finite, nonnegative value $v_i(S)$ to each piece $S$.
Values add over disjoint pieces. Individual points have value zero, and
the value of $[a,x]$ changes continuously as $x$ moves from $a$ to $1$.
Agents are \emph{hungry}: $v_i([a,b])>0$ whenever $a<b$.
Thus this cumulative value is strictly increasing as well as continuous.
These two facts are what make prescribed-value cuts possible and unique.
The precise measure assumptions are recorded in
\cref{app:measure-conventions}; no density or smoothness assumption is
needed. Total values need not equal one and are initially unknown.

Endpoints have zero value to everyone. We therefore treat pieces
differing at finitely many endpoints as the same, and assign any
unallocated endpoints arbitrarily when returning a complete division.
This uses no queries and changes no values. Throughout, a piece is called
\emph{empty} if it contains no interval of positive length; it may still
contain finitely many endpoints.

\begin{definition}[Allocations and envy-freeness]\label{def:allocation}
An allocation $X=(X_i)_{i\in A}$ of a piece $U$ among agents $A$
consists of disjoint pieces $X_i\subseteq U$, called their \emph{bundles}.
The cake left unallocated is the \emph{residue}
$T=U\setminus\bigcup_iX_i$. The allocation is \emph{complete} if this
residue is empty, up to endpoints. It is \emph{envy-free} if
\[
 v_i(X_i)\ge v_i(X_j)\qquad(i,j\in A).
\]
Both sides are evaluated by agent $i$: she must prefer her own bundle
to every other bundle, or be indifferent.
\end{definition}

Allocations may be partial. When $X$ and $Y$ allocate disjoint subcakes,
$X\oplus Y$ gives each agent her pieces from both allocations; an agent
absent from one receives nothing from it.

\subsection{Queries, randomness and cost}

We learn values by asking individual agents two kinds of question:
the value of an interval, or a cut giving a specified value.

\begin{definition}[Value and cut queries]\label{def:protocol}
An evaluation query $\Eval_i(a,b)$, with $0\le a\le b\le1$, returns
$v_i([a,b])$. A cut query $\Cut_i(a,t)$, with
$0\le t\le v_i([a,1])$, returns the unique $x\in[a,1]$ such that
$v_i([a,x])=t$.
\end{definition}

For example, to take half an agent's value of $[a,b]$, first ask its
value and then cut from $a$ at half that value. The answer need not be
the midpoint: value and length are different. Every cut requested by
the protocol has a feasible target; its construction checks this from
values already known. Queries concern an individual agent's valuation.
A cut with respect to an average or a difference of valuations is not
an additional allowed query and must be implemented using these two
types.

The protocol draws independent uniform real numbers in $(0,1)$ as
needed. Valuations remain fixed; probabilities and expectations concern
only these draws. A bound \emph{given the past}, or
conditional on the \emph{history}, fixes all preceding queries, answers
and random choices. The next draw is independent of that history.
A bound \emph{for every run} holds for every sequence of draws.
The formal conventions ensuring that these probabilities are well
defined appear in \cref{app:measure-conventions}.

Query complexity counts questions to agents, but not calculations with
their answers~\cite[Section~2]{Procaccia2009}. Trying every assignment of
known bundles, for example, needs no new queries but may require too
much computation. We therefore count computation separately.

\begin{definition}[Computation cost]\label{def:computation-model}
Each query, independent uniform-real draw, exact arithmetic operation
or comparison counts as one operation.
\end{definition}

This is a bound for exact real arithmetic, not binary running time.

\subsection{Local calls and recorded values}

The protocol repeatedly divides smaller pieces among smaller groups.
Write $N=n$ for the original number of agents, $A$ for the agents in a
local call and $k=|A|$ for their number. The original population $N$
stays fixed, while $k$ may decrease. Valuations in a local call are the
original valuations restricted to its input cake; any rescaling will be
stated explicitly.

To decide when agents can keep their shares and leave, we compare their
advantages with the cake still unallocated. The \emph{envy margin}, or
\emph{advantage}, of $i$ over $j$ is
\[
 \Delta_{ij}(X)=v_i(X_i)-v_i(X_j).
\]
In an envy-free allocation this is the value, measured by $i$, that
bundle $X_j$ can gain before $i$ would prefer it to her own.
Agent $i$ \emph{dominates} $j$ if
$\Delta_{ij}(X)\ge v_i(T)$: her advantage covers her value of the whole
residue. She would then not envy $j$ even if $j$ received all of it.
This is also called irrevocable advantage~\cite{BramsTaylor1995}.
An agent can keep her bundle and stop taking part once she dominates
everyone who will divide the residue (\cref{sec:completion}).

Our pieces may have several components, whereas a query concerns a
single interval. We bridge this difference by recording every agent's
value of the cake up to each endpoint. Later values of pieces then need
only arithmetic.

\begin{definition}[Ordered pieces and recorded endpoint values]\label{def:cache}
An \emph{ordered finite union} has its disjoint interval components
listed in a fixed order. Traversing them in that order, each from left
to right, and skipping the gaps defines its \emph{virtual order}.
A consecutive segment in this traversal is a \emph{logical cell}; it
may consist of several intervals in the cake.

Let $\mathcal K$ be the endpoints in the current working table. For every
$x\in\mathcal K$ and every original agent $j$, we record the prefix value
$v_j([0,x])$. Initially $\mathcal K=\{0\}$, with recorded value zero.
A piece has \emph{recorded endpoint values} when all its endpoints belong
to $\mathcal K$. Recording a new point $x$ requires at most $N$
evaluation queries $\Eval_j(0,x)$. In particular, recording $1$ costs $N$
queries, and recording all endpoints of an input with $C$ components costs
at most $2NC$ queries.
\end{definition}

The virtual order reads one component from left to right, then the next,
skipping the gap. A logical cell can finish one component and continue
in the next; this is why it need not be a single physical interval.

For example, let $S=[a,b]\cup[c,d]$, where
$0\le a\le b\le c\le d\le1$. Once the four endpoint values are
recorded, agent $j$'s value of $S$ is
\[
 v_j(S)=\bigl(v_j([0,b])-v_j([0,a])\bigr)
       +\bigl(v_j([0,d])-v_j([0,c])\bigr).
\]
No further queries are needed. The same subtraction and addition gives
every agent's value of any piece with recorded endpoint values.

A \emph{prefix} of an ordered finite union is an initial segment in its
virtual order. To cut one of a prescribed value, locate the target
component using recorded values, cut within it and record the new
endpoint for all agents: at most one $\Cut$ and $N$ $\Eval$ queries.
Here the prefix follows the chosen component order; the recorded prefix
values in \cref{def:cache} always start at $0$ in the whole cake.
If the target is reached exactly at a component
endpoint, reuse it without a cut. Gaps lie outside this piece and are
skipped in the traversal
(\cref{lem:union-cut}).

\begin{remark}[Recording values for every original agent]\label{rem:cache-invariant}
Every new endpoint is immediately recorded for all $N$ original agents,
including agents absent from a local call. Thus every bundle and
residue has recorded endpoint values, including the outputs of
\cref{thm:main,thm:partial}, capped or uncapped. All exact checks
use values determined by query answers.
\end{remark}

This is why local query bounds involve both $N$ and $k$. Combining
pieces by unions, differences or reassignment introduces no endpoints.
Only cuts increase the number of intervals, each by a bounded amount
(\cref{lem:additive-growth}).

\paragraph{Verified outputs and retries.}
\Cref{sec:overview} gives the guarantees of our three procedures:
Repair, Shrink and Snapshot. They make tentative choices in
\emph{trials}. Before accepting an output, they check disjointness,
identify the unallocated cake and verify every promised inequality
exactly from recorded values. Rejected trials allocate nothing.
Every local guarantee applies to every permitted input and preceding
history, including cake chosen using earlier random outcomes.
Its query bound includes recording new endpoints for all original
agents. The retry bound in \cref{lem:retry} proves termination and
bounds cost whenever each trial has a stated positive chance of success.

An outer trial may call Repair once inside it. For counting retries,
an \emph{attempt} is the entire outer trial: one snapshot trial, one
shrinking step, or one standalone repair trial.
It includes preparation, at most one inner repair and the final
acceptance checks. A failed inner repair ends its entire outer attempt;
even a successful inner repair is temporary until those final checks
pass.

A failed attempt restores its saved working state, discarding new
tables, endpoint records and tentative pieces. Previously successful
work and all cost counters are retained: discarded queries and
operations still count.
Fresh draws are independent of the full past, including discarded
observations. Assignments remain provisional until the protocol
returns; the capped version may instead report failure
(\cref{sec:accounting}).

Working records and a trial's sampling grid have different roles.
Later we estimate values inside cells by straight lines between recorded
boundaries (\cref{def:sampling-model}). Recording another endpoint makes
its values available for exact checks; it does not change this fixed
cell estimate. This applies to inherited records and to new endpoints
generated during the trial.

\paragraph{Notation.}
The original population is $N$ and the current group size is $k$.
The residue $T$ is cake still unallocated; $\Delta_{ij}(X)$ measures
agent $i$'s advantage over agent $j$.
We write $[k]=\{1,\ldots,k\}$, use natural logarithms unless specified,
and write $\osc_jx_j=\max_jx_j-\min_jx_j$ for the range of a finite list.
\section{Local guarantees}\label{sec:overview}

An envy-free allocation may leave cake over. Making the remainder
smaller is useful, but does not by itself finish the division: a positive
amount may remain after every finite number of steps. We instead give
some agents advantages that cover everything the others could receive
from the remainder. Those agents can keep their shares and stop.

Three local procedures supply what this argument needs. \emph{Repair}
removes small initial envy while keeping old bundles whole and using
little extra cake. \emph{Shrink} allocates enough cake to reduce every
agent's value of the remainder by a prescribed factor. \emph{Snapshot}
produces an allocation with an advantage measured against the cake it
leaves unallocated. We state these guarantees here and use them to complete
the division in \cref{sec:completion}; their constructions follow in
\cref{sec:kernel,sec:snapshot}.

\begin{figure}[!htb]
\centering
\begin{tikzpicture}[
  >={Latex[length=2mm,width=1.4mm]},
  box/.style={draw,rounded corners=2pt,align=center,text width=6.8cm,
    inner sep=6pt,font=\normalsize,execute at begin node=\hyphenpenalty10000},
  stage/.style={box,draw=aliceblue!85!black,fill=aliceblue!6},
  exit/.style={box,text width=4.6cm,draw=carolgreen!75!black,
    fill=carolgreen!6},
  flow/.style={->,line width=.7pt}]
\node[box,draw=residuegray,fill=residuegray!6] (entry) at (0,0)
 {\textbf{Current agents; remaining cake}};
\node[stage,below=.65cm of entry] (save)
 {\textbf{Save useful partial allocations}\\
  Keep bundles with advantages over others.};
\node[stage,below=.4cm of save] (shrink)
 {\textbf{Shrink the remaining cake}\\
  The saved advantages become large\\relative to what remains.};
\node[stage,below=.4cm of shrink] (reassign)
 {\textbf{Reassign saved bundles}\\
  Each agent loses little; a group gains\\advantages over those outside it.};
\node[stage,below=.4cm of reassign] (repair)
 {\textbf{Repair all current agents}\\
  Remove envy while preserving\\the group's advantages.};
\node[box,draw=boborange!85!black,fill=boborange!8,below=.4cm of repair] (fixed)
 {\textbf{Fix one group's shares}\\
  Its advantages cover everything\\the others could still receive.};
\draw[flow] (entry.south) -- (save.north)
  node[midway,right,font=\small] {four or more};
\draw[flow] (save.south) -- (shrink.north);
\draw[flow] (shrink.south) -- (reassign.north);
\draw[flow] (reassign.south) -- (repair.north);
\draw[flow] (repair.south) -- (fixed.north);
\coordinate (loop) at ($(entry.west)+(-.65,0)$);
\coordinate (loop-top) at ($(entry.north)+(0,.55)$);
\draw[flow] (fixed.west) -- (loop |- fixed.west)
  -- (loop |- loop-top) -- (loop-top) -- (entry.north);
\node[anchor=south west,font=\small,inner sep=3pt]
 at (loop |- loop-top)
 {Fewer agents; remaining cake};

\node[exit,right=1.5cm of entry] (base)
 {\textbf{At most three agents}\\Complete directly.};
\node[exit,anchor=north] (finish) at (base.center |- repair.north)
 {\textbf{Complete allocation}\\
  Check coverage and envy,\\then return all shares.};
\draw[flow] (entry.east) -- (base.west);
\coordinate (outer) at ($(base.east)+(.45,0)$);
\draw[flow] (base.east) -- (outer) -- (outer |- finish.east) -- (finish.east);
\coordinate (empty) at ($(entry.east)+(.65,0)$);
\draw (entry.east) -- (empty) -- (empty |- shrink.east);
\draw (save.east) -- (empty |- save.east);
\draw (shrink.east) -- (empty |- shrink.east);
\coordinate (exit-turn) at ($(reassign.south)+(0,-.2)$);
\draw[flow] (empty |- shrink.east) -- (empty |- exit-turn)
  -- (finish.north |- exit-turn) -- (finish.north);
\node[anchor=west,align=left,text width=4.2cm,font=\small,inner sep=4pt]
 at ($(empty |- shrink.center)+(.15,0)$)
 {No cake remains:\\exit before saving, or\\after saving or shrinking.};
\node[align=center,text width=4.6cm,font=\normalsize]
 at (base.center |- fixed.center) {Earlier fixed shares\\stay with their owners.};
\end{tikzpicture}
\caption{The completion loop. Allocations are saved on successive
remainders. After reassignment and repair, one group can stop because
its advantages cover all the cake left to divide. The others keep their
allocated bundles and repeat on the remaining cake. Arrows show successful
progress; failed local trials retry. Repair is also used inside the
saving and shrinking procedures.}
\label{fig:protocol-overview}
\end{figure}
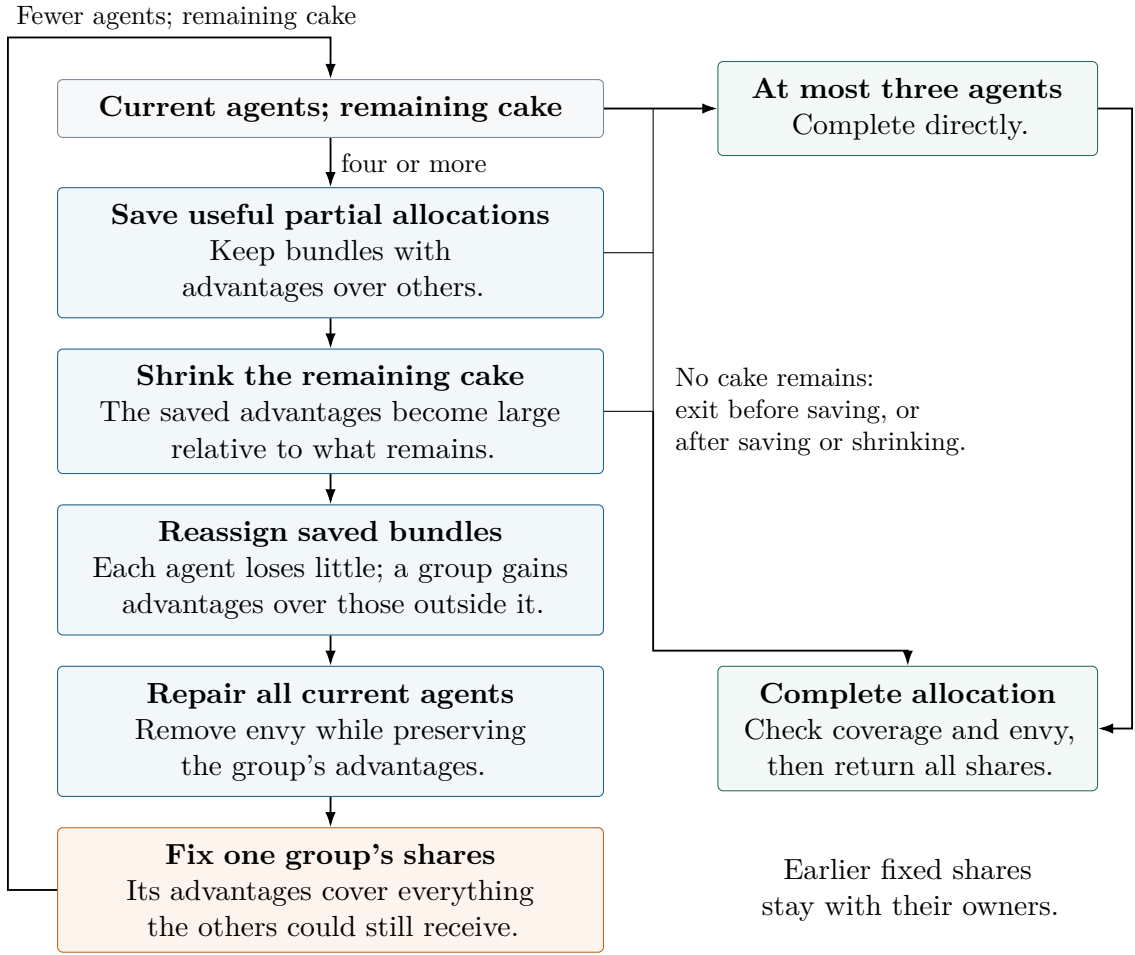

\subsection{Whole-bundle repair}

Suppose each agent already has a bundle, and some additional cake is
available to remove envy. We call the old bundles \emph{endowments} and
the additional cake the \emph{buffer}. An endowment may change hands,
but it must not be split. Repair is possible when any agent's preference
for another endowment is small compared with her value of the buffer.
She may prefer her own endowment by a large amount; only envy needs a
bound.

Formally, there are $k\ge2$ participants among the $N$
original agents, and disjoint finite unions $B,W_1,\ldots,W_k$ with
recorded endpoint values: $B$ is the available cake and the $W_j$ are the
endowments. Initially $W_i$ belongs to agent $i$. Write $d_i=v_i(B)$.
The input condition bounds each agent's envy by a fraction $\tau$ of
her buffer value:
\begin{equation}\label{eq:kernel-input}
 v_i(W_j)-v_i(W_i)\le\tau d_i\qquad(i,j\in[k]).
\end{equation}
If the buffer is empty, apart from possible endpoints, this says that
the input assignment is already envy-free. Return it unchanged, with
no additions. Otherwise the buffer contains an interval of positive
length, so hungriness gives $d_i>0$ for every participant. When all
endowments are empty, the condition holds automatically.

All numerical choices use the fixed base
\[
 s=2^{20}\max\{4,N\}.
\]
This base depends only on the original population $N$ and stays fixed
when the participating group becomes smaller. It is not a parameter
chosen from the agents' valuations. The fraction
$\tau=1/(4s^{13})$ is the \emph{repair tolerance}; the remaining numerical
choices are collected in \cref{def:kernel-parameters}.

\Needspace{12\baselineskip}
\begin{theorem}[Partial allocation and whole-bundle repair]\label{thm:kernel}\label{lem:kernel-zero}
Under \eqref{eq:kernel-input}, one trial of $\operatorname{Repair}$ either rejects without
allocating cake, or returns a permutation $\pi$ and disjoint finite-union additions
$Y_i\subseteq B$. The allocation with bundles
\[
 X_i=W_{\pi(i)}\cup Y_i
\]
is envy-free and gives every agent at least her value of every input
endowment: $v_i(X_i)\ge\max_jv_i(W_j)$.
Every endowment is assigned whole.
The additions together use at most a quarter of the buffer in every
participant's valuation:
\[
 v_i\Bigl(\bigcup_jY_j\Bigr)\le\tfrac14v_i(B)\qquad(i\in[k]).
\]
Every trial uses $O(Ns^{24})$ queries. Conditional on every history at
its start, it returns with probability above $0.98$.
\end{theorem}

The value-preservation guarantee follows from whole bundles and
envy-freeness: every old endowment lies in someone's final share, which
agent $i$ cannot prefer to her own. The expenditure guarantee does a
different job. If she prefers one old bundle to another by more than a
quarter of her buffer value, so little extra cake cannot compensate her
for receiving the worse one.
This observation is the last step of the completion proof. The repair
construction and its probability analysis are in \cref{sec:kernel}.

\subsection{Smaller remainders and useful advantages}

Repair also lets us allocate almost all of the cake from scratch.
Carve endowments that are nearly equal in every agent's valuation,
leaving a small buffer. The carving is accurate enough that this buffer
can still repair their discrepancies. Once the endowments are allocated,
the residue is worth at most $s^{-8}$ of the input to every agent.
Repeating gives the next guarantee, proved in \cref{sec:kernel-shrink}.

\begin{lemma}[Repeated shrinking]\label{lem:shrink}
For $k$ agents, a finite union $T$ with recorded endpoint values and
$G\ge1$, the procedure $\operatorname{Shrink}(G)$ terminates almost surely
and returns an envy-free partial allocation $Y$ of $T$, leaving $T'$ with
$v_i(T')\le v_i(T)/G$ for every participating agent. Adding $Y$ to an
allocation disjoint from $T$ cannot decrease any existing envy margin
between the participating agents. For $k\ge2$, its expected query cost is
\[
 O\!\left(Ns^{24}(1+\log G/\log s)\right).
\]
Each shrinking attempt uses $O(Ns^{24})$ queries, and the expected number
of attempts is $O(1+\log G/\log s)$, conditional on any history before the
procedure starts. For $k=1$, give that agent all of $T$.
\end{lemma}

Shrinking preserves existing advantages, but we still need an advantage
to start with. A gap of $10^{-100}$ is not useful protection if the
remainder is worth $1/2$. The next procedure gives a gap that is an
inverse-polynomial fraction of its owner's remaining-cake value. We
save its allocation and a pair of agents $a,b$ recording the gap: agent
$a$ prefers her share to agent $b$'s. Such a saved allocation is called a
\emph{snapshot}, and $(a,b)$ is its \emph{certificate pair}.

\begin{theorem}[A margin relative to the residue]\label{thm:snapshot}
For $k\ge1$ agents and a finite union $U$ with recorded endpoint values,
the procedure $\operatorname{Snapshot}$ terminates almost surely and
returns either a complete envy-free allocation, or an envy-free partial
allocation $X$ with residue $T$ and distinct agents $a,b$ such that
\begin{equation}\label{eq:snapshot-certificate}
 v_a(X_a)-v_a(X_b)\ge\theta v_a(T),
\end{equation}
where $m=s^{24}$ is the sample count and $\theta=\tau/(768m)$
is inverse-polynomial in the original agent count $N$. Every output is checked
exactly, and all unallocated cake stays in $T=U\setminus\bigcup_hX_h$.
Each attempt uses $O(Ns^{24})$ queries and succeeds with probability
at least $1/(64k)$, conditional on any history at its start. Repeating rejected
attempts gives expected query cost $O(Nks^{24})$.
\end{theorem}

One snapshot gives an advantage over just one other agent. Completion
needs a group whose members have advantages over everyone outside it.
We save several snapshots and shrink their common residue until each
saved advantage covers twice its owner's value of that residue.
Reassigning selected bundles at small loss produces the group we need.
One repair among all active agents then removes the envy caused by
reassignment while preserving enough advantage for that group to stop
(\cref{sec:completion}). No local guarantee assumes the completion theorem;
the proofs of the local procedures use only queries, finite calculations
and sampling. The construction and cost of Snapshot are proved in
\cref{sec:snapshot,prop:snapshot-cost}.

\subsection{Numerical credits and cake payments}

The new local ingredient is compensation with actual cake. Repair sets
aside a \emph{reserve}, a portion of its buffer used for payments, and
checks that every agent values it positively. Divide each agent's values
by her own reserve value; this positive rescaling does not change her
preferences. The reserve is then worth one to everyone. A \emph{credit}
attached to a bundle is an ideal payment: every agent adds the same
number when evaluating that bundle. A physical cake payment need not
have the same value to everyone, so we must control the difference.

\begin{figure}[!htb]
\centering
\begin{tikzpicture}[>=Latex, every node/.style={font=\normalsize}]
\node[draw=boborange!80!black,fill=boborange!4, rounded corners=2pt, inner sep=7pt] (before) at (0,0) {
\begin{tabular}{lcc}
 & Alice's & Bob's\\[-1pt]
 & bundle & bundle\\\hline
Alice & $\mathbf{0.60}$ & $0.50$\\
Bob   & $0.40$ & $\mathbf{0.32}$
\end{tabular}};
\node[draw=carolgreen!75!black,fill=carolgreen!5, rounded corners=2pt, inner sep=7pt] (after) at (8.0,0) {
\begin{tabular}{lcc}
 & Alice's & Bob's\\[-1pt]
 & bundle & bundle\\\hline
Alice & $\mathbf{0.62}$ & $0.61$\\
Bob   & $0.42$ & $\mathbf{0.43}$
\end{tabular}};
\draw[->,thick] (before.east) -- (after.west)
 node[midway,above,align=center] {Ideal credits\\Alice: $+0.02$\\Bob: $+0.11$};
\node[below=7pt of before,align=center] {Bob prefers Alice's bundle};
\node[below=7pt of after,align=center] {Both prefer their own bundles};
\end{tikzpicture}
\caption{Illustrative credits, not a protocol run. Compare entries within
each row: the row names the agent evaluating the two bundles, and bold
entries are her own bundle's value. Values are in units of that agent's
reserve value. Add $0.02$ to both entries in Alice's bundle column and
$0.11$ to both entries in Bob's: each agent then has margin $0.01$.
If cake pays each recipient exactly and changes the other entry in
each row by at most $0.004$, both margins remain at least $0.006$.}
\label{fig:credit-example}
\end{figure}
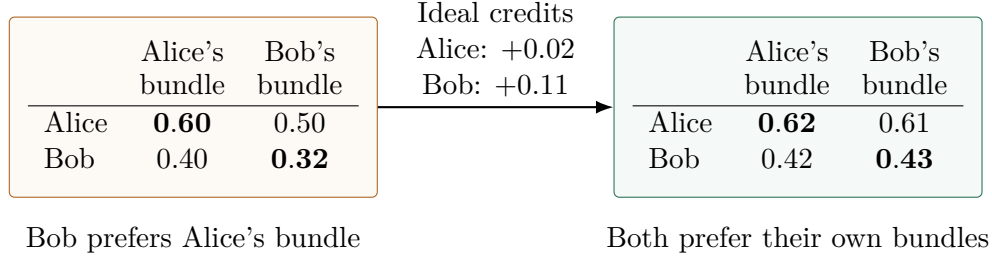

In \cref{fig:credit-example}, Bob initially envies Alice. Credits $0.02$
and $0.11$ remove his envy while leaving Alice satisfied. Both final
advantages are $0.01$, so the ideal payments have room for small errors.

For fixed bundles, Halpern and Shah's subsidy construction gives
nonnegative credits after an assignment maximising total value
\cite[Theorem~1]{HalpernShah2019}. Bounds on subsidies are also studied
for indivisible goods~\cite{BrustleEtAl2020} and mixed goods and
chores~\cite{LuMackenzieSuzuki2026}. Using cake to realise payments
appears in Aziz, Lu, Mackenzie and Suzuki's mixed-resource construction
\cite[Section~3]{AzizLuMackenzieSuzuki2025}. Here we need room for
payment errors as well as the removal of envy.

The buffer is divided into small pieces, called \emph{cells}, whose
values are recorded for every agent. A fraction of a cell's recorded
value specifies a numerical payment; it does not specify a physical
piece with that fraction of everyone's value. We adjust these fractions
while preserving all payment totals until only a few cells are split.
Giving the other cells whole leaves small deficits. We fill each deficit
from the unused reserve and check the completed payment: its recipient
must value it at exactly her credit, and every other agent's evaluation
must stay within the permitted error. The total credits and permitted
errors use only a small part of the reserve, leaving enough for each
recipient in turn.

Randomness makes the necessary margins and errors comparable. A random
selection of cells gives a reserve worth roughly the same fraction of
the buffer to every agent, with high probability. Other random choices
make near ties between competing assignments unlikely and treat the old
endowments symmetrically.
Independent sample points record more of the valuations before payments
are cut. A fresh starting point for each correction, uniform in its
recipient's value, then controls how differently the agents value it. The
construction and its exact checks are given in \cref{sec:kernel}.

Both the margins and the allowed errors shrink with the agents'
recorded disagreement. Nearly agreeing valuations therefore do not
force a larger query bound. The probability estimates average over
samples and later cuts together; they do not promise success for every
observed table.

The choices made from a recorded table use no new queries. Ye and Bai
use conditional expectations to make assignments of fully
evaluated pieces deterministic~\cite[Section~5]{YeBai2026}. New cut
points, however, require new valuation information. Knowing that some
random choices work does not provide a way to find them with few
queries; a deterministic polynomial query bound remains open.
\Needspace{10\baselineskip}
\section{Complete envy-free allocations}\label{sec:completion}

We first show how the local guarantees from the previous section finish
an allocation; their constructions follow in the next two sections.
A useful partial allocation gives one agent an advantage measured
against the cake left over. Completion needs more: a group whose
members have enough advantage over everyone outside the group to
cover all the cake those agents might receive later.
Recall that $i$ \emph{dominates} $j$ when her margin over $j$
is at least her value of the residue. If every member of a nonempty
group $S$ dominates everyone outside it, members of $S$ can keep their
bundles and stop. Only the other agents need divide the residue.

The saved allocations, separated margins, permitted cycles and
domination framework come from Ye and Bai~\cite[Sections~4.3--4.6]{YeBai2026}.
Their completion uses the Core and SubCore procedures introduced by
Aziz and Mackenzie~\cite{AzizMackenzie2016}. Here reassignments at small
loss identify two groups, and one whole-bundle repair preserves enough
of the first group's advantages to let it stop.

We use the snapshot procedure of \cref{thm:snapshot}, the shrinking
procedure of \cref{lem:shrink}, and whole-bundle repair from
\cref{thm:kernel}, repeated until it succeeds. Each guarantee holds
after every preceding history. All other choices use fixed rules on
recorded values.

\Needspace{8\baselineskip}
\begin{theorem}[Completion]\label{thm:completion}
Repair, Shrink and Snapshot yield a protocol that completely and envy-freely
allocates any finite union $C$ with recorded endpoint values among the
$N$ agents. For every hungry valuation profile and preceding history,
it terminates almost surely and checks the output exactly
before returning it. Randomness occurs only inside the three procedures.
\end{theorem}

One \emph{stage} has $k\ge4$ active agents $A$ and input cake $C$;
at most three active agents use \cref{lem:base-cases}. It either completes the
allocation or lets a nonempty group stop and passes the residue to a
smaller group. Only pieces allocated during this stage may be
reassigned. All bundles fixed in earlier stages stay with their owners.
The stage has the following order.
\begin{enumerate}
\item Save $k^2$ snapshots on successive residues. If one finishes the division,
combine it with the earlier allocations and finish the stage.
\item Shrink the remaining cake until each saved comparison is either
a small reassignment loss or a large advantage. Keep the allocations
made by shrinking with their original recipients.
\item Reassign saved bundles so that each agent loses little relative to her
original bundles, while some nonempty group $S$ has a large advantage
over everyone outside it.
\item Repair all active agents together, repeating failed repair attempts
until one returns. Members of $S$ then dominate those outside $S$:
they keep their resulting bundles and stop, and the others divide the
remaining cake recursively.
\end{enumerate}
Before collection and after every snapshot or shrinking call, an empty
residue finishes the stage with all existing assignments retained.
This includes a residue consisting only of
endpoints, which can be assigned without changing anyone's value.
Otherwise the residue contains an interval of positive length and is
worth something to every agent, by hungriness. We give the steps and
their proofs together, then count the calls in
\cref{prop:completion-stage}.

Two elementary facts let us combine the stage's allocations. On
disjoint cake, envy margins add:
\begin{equation}\label{eq:completion-addition}
 \Delta_{ij}(X\oplus Y)=\Delta_{ij}(X)+\Delta_{ij}(Y).
\end{equation}
Thus adding an envy-free allocation preserves earlier margins.
This fact and the stopping rule above follow
\cite[Lemma~4.1 and Corollary~4.2]{YeBai2026}.

\subsection{Saved comparisons}\label{sec:completion-procedure}

\paragraph{Snapshots and a common residue.}
In saved snapshot $t$, write $C_{t,j}$ for the bundle originally
assigned to $j$. Its \emph{label} $j$ records that original recipient
and stays with the bundle when it changes hands. Record the margins
\[
 d_{t,i,j}=v_i(C_{t,i})-v_i(C_{t,j})\ge0.
\]
This is $i$'s advantage over label $j$ in that snapshot; it is
nonnegative because the snapshot is envy-free. Save also its
certificate pair $(a_t,b_t)$ and the residue $T_t$ it leaves, so that
\eqref{eq:snapshot-certificate} gives
$d_{t,a_t,b_t}\ge\theta v_{a_t}(T_t)$.

Let $T$ be the residue after collecting the snapshots. Apply
$\operatorname{Shrink}(2/\theta)$, where $\theta=\tau/(768m)$ is the
snapshot guarantee. Keep the allocation made by this and every later
shrinking call with its original recipients; write $H$ for their sum.
If cake remains, the new residue $T'$ satisfies
\[
 2v_{a_t}(T')\le\theta v_{a_t}(T)
 \le\theta v_{a_t}(T_t)\le d_{t,a_t,b_t}
 \qquad\text{for every saved }t.
\]
Here $T\subseteq T_t$, since later snapshots use only cake left by
earlier ones. Each saved allocation now has at least one advantage
worth twice its owner's remaining cake. This factor of two leaves
room for the final repair; further shrinking preserves it.

\paragraph{Small losses and large advantages.}
Each saved margin is measured against $i$'s original bundle. Receiving
label $j$ costs $i$ exactly $d_{t,i,j}$; giving it to another agent
provides that much advantage before subtracting $i$'s own reassignment
loss.
We want every comparison to be either a small loss or a large
advantage. Following the margin-separation idea of
Ye and Bai~\cite[Lemma~4.7]{YeBai2026}, we keep the bundles fixed and
shrink only the residue until every comparison has one of these uses.

The repair tolerance $\tau$ is our entire allowance for reassignment
losses. Dividing it across the $k^2$ snapshots limits their total loss;
an advantage worth twice the residue will leave room for that loss and
the final repair. For the current residue $T$, call
a margin of agent $i$ \emph{small} if it is at most
$\tau v_i(T)/k^2$, and \emph{large} if it is at least $2v_i(T)$.
Whenever a saved margin $d$ lies strictly between these thresholds,
apply $\operatorname{Shrink}(2k^2/\tau)$, adding its allocation to $H$.
The new residue satisfies
\[
 2v_i(T')\le\frac{\tau}{k^2}v_i(T)<d,
\]
so the triggering margin is now large and stays large. Other small
margins may become intermediate later, but each can trigger only once.
There are $k^2$ snapshots with $k(k-1)$ nontrivial comparisons each,
so fewer than $k^4$ shrinking calls suffice.

If cake remains, fix the resulting residue $T_*$ and write
$\rho_i=v_i(T_*)>0$. These reference values stay fixed below.
An agent \emph{marks} label $j$ in snapshot $t$ when
$d_{t,i,j}\ge2\rho_i$.
An unmarked label has margin at most $\tau\rho_i/k^2$. No agent
marks her own label. Every snapshot has some mark, because its
recorded comparison remained large during shrinking.

\subsection{Two groups of agents}

We now transfer saved bundles so that one group has a large advantage
over every member of the other, while each recipient loses little.
For snapshot $t$, form a \emph{permitted graph} $P_t$: its edge $h\to j$
means that $h$ leaves label $j$ unmarked, and can therefore receive
$C_{t,j}$ at a loss of at most $\tau\rho_h/k^2$. Every self-loop is
present, and at least one edge is missing because the snapshot has a
large recorded comparison. This is the permitted graph of
Ye and Bai~\cite[Section~4.5]{YeBai2026}.

A cycle of permitted edges describes a safe exchange: each agent on
the cycle receives the bundle labelled by her successor, and everyone
else keeps her own. Write $\pi_t(h)$ for the label received by $h$.
The graph lemma chooses exchanges with two properties: each agent
receives an unmarked label in every snapshot, keeping her total loss
small; and for each $i\in S$, $h\notin S$, some snapshot gives $h$ a
label marked by $i$, supplying their large comparison.

\begin{lemma}[Permitted transfers]\label{lem:completion-graph}
Let $P_1,\ldots,P_{k^2}$ be directed graphs on the same $k\ge2$
vertices. Each contains every self-loop and misses at least one edge.
There are permutations $\pi_t$ and a nonempty proper set $S$ such that
$h\to\pi_t(h)$ is an edge of $P_t$ for every $h,t$, and, for every
$i\in S$, $h\notin S$, some $t$ has $i\to\pi_t(h)$ absent from $P_t$.
\end{lemma}
\begin{proof}
A directed graph is \emph{strongly connected} if every vertex can
reach every other. If one $P_t$ is not strongly connected, take a
\emph{sink component} $S$: a nonempty proper group whose vertices can
reach one another and which has no edge leaving it. Keeping every
permutation equal to the identity proves the claim in this case;
compare~\cite[Lemma~4.10]{YeBai2026}.

Otherwise every $P_t$ is strongly connected. Start a second graph $G$
with every edge between distinct vertices. An edge $i\to h$ in $G$
records an unfinished task: find a snapshot in which $h$ receives a
label marked by $i$. While $G$ is strongly
connected, take a fresh $P_t$ and a missing edge $a\to j$ in it.
On a path in $G$ from $a$ to $j$, the first vertex has no permitted
edge to $j$, whereas the last has its self-loop. Some path edge
$i\to h$ therefore crosses from a vertex without such an edge to
one with it. Thus $i\to j$ is absent from $P_t$
and $h\to j$ is present.

The edge $h\to j$ lies on a simple directed cycle in $P_t$. Rotating its
labels gives a permitted permutation with $\pi_t(h)=j$; if $h=j$,
use the identity. This is the permitted-cycle exchange
of~\cite[Lemma~4.11]{YeBai2026}. Delete $i\to h$ from $G$ and never
change this permutation again. Each step supplies one comparison and
deletes a distinct edge. There are fewer than $k^2$ edges, so $G$
ceases to be strongly connected before the graphs run out. Take a
sink component $S$ of the final $G$ and use identity permutations for
unused snapshots. No edge of the final $G$ leaves $S$. Since $G$ initially contained
every edge between distinct vertices, every edge from $S$ to its
complement was deleted; its stored permutation supplies the required
missing edge.
\end{proof}

Apply the lemma to the permitted graphs and put
$\bar S=A\setminus S$. Combine each agent's reassigned snapshot
bundles with her unchanged shrinking allocation:
\[
 D_h=H_h\cup\bigcup_t C_{t,\pi_t(h)}\qquad(h\in A).
\]
These combined bundles are the endowments for the final repair;
$D_h$ now carries label $h$.

\subsection{One global repair}

The combined bundles cover the stage's input outside $T_*$, are disjoint
and have recorded endpoints. We show that they meet repair's input
condition and that their cross-group advantage survives repair. Use
$T_*$ as the buffer and repair all active agents together, repeating
attempts until one returns.

Two repair guarantees matter here. Preserving each agent's own
endowment value prevents a member of $S$ from receiving an endowment
labelled outside $S$. Limiting the extra cake to a quarter of the
buffer's value then preserves enough advantage for $S$ to stop.

\begin{lemma}[Global repair and domination]\label{lem:completion-global-repair}
Every returned repair with endowments $(D_i)_{i\in A}$ and buffer $T_*$
is an envy-free allocation $Z$. It assigns the $S$-labelled endowments to
$S$ and the $\bar S$-labelled endowments to $\bar S$, and satisfies
\begin{equation}\label{eq:completion-cross-margin}
 \Delta_{ih}(Z)\ge(7/4-\tau)\rho_i>\rho_i
 \qquad(i\in S,\ h\in\bar S).
\end{equation}
Every member of $S$ therefore dominates every member of $\bar S$.
\end{lemma}
\begin{proof}
Each reassignment subtracts the recipient's own loss from her advantage
against the other recipient's new label. Thus
\[
 \Delta_{ih}(D)=\Delta_{ih}(H)+
 \sum_t\bigl(d_{t,i,\pi_t(h)}-d_{t,i,\pi_t(i)}\bigr).
\]
Shrinking is envy-free, every first margin in the sum is nonnegative,
and every subtracted margin is at most $\tau\rho_i/k^2$. Across the
$k^2$ snapshots the total loss is therefore at most $\tau\rho_i$.
For $i\in S$, $h\in\bar S$, the graph lemma supplies one first margin
of at least $2\rho_i$. Consequently
\begin{align}
 v_i(D_i)-v_i(D_h)&\ge-\tau\rho_i &&(i,h\in A),
       \label{eq:completion-small-envy}\\
 v_i(D_i)-v_i(D_h)&\ge(2-\tau)\rho_i
       &&(i\in S,\ h\in\bar S).
       \label{eq:completion-witness}
\end{align}
The first bound validates repair with buffer value $\rho_i>0$.
Write its return as $Z_i=D_{\pi(i)}\cup Y_i$, where $\pi$ permutes
whole endowments and $Y_i\subseteq T_*$. Envy-freeness implies
$v_i(Z_i)\ge v_i(D_i)$, since $D_i$ lies in someone's final bundle.
Repair also gives
$v_i(\bigcup_hY_h)\le\rho_i/4$.
If $i\in S$ received an endowment labelled in $\bar S$, then
\eqref{eq:completion-witness} would give
\[
 v_i(Z_i)\le v_i(D_i)-(2-\tau)\rho_i+\rho_i/4<v_i(D_i),
\]
contradicting value preservation. Every member of $S$ therefore receives
an $S$-labelled endowment. Since the assignment is a permutation,
every member of $\bar S$ receives a $\bar S$-labelled one.

For $i\in S$ and $h\in\bar S$, the label $j=\pi(h)$ lies outside
$S$. Value preservation and the bound on additions give
\[
 v_i(Z_i)-v_i(Z_h)
 \ge v_i(D_i)-v_i(D_j)-\rho_i/4
 \ge(7/4-\tau)\rho_i>\rho_i.
\]
Every unallocated point lies in $T_*$, worth $\rho_i$ to $i$.
Thus the displayed margin already covers the residue. All reverse
comparisons, and those within either group, follow directly from the
envy-freeness of the global repair.
\end{proof}

\subsection{Domination and recursive completion}\label{sec:completion-cost}

Members of $S$ keep their bundles and stop; only $\bar S$ divides the
residue. Agents in $\bar S$ already do not envy $S$, and their additional
cake cannot create such envy. Agents in $S$ can tolerate even the
whole residue going to one member of $\bar S$. The recursive problem thus
has fewer agents, and requires no further shrinking before it starts.
Bundles fixed in earlier stages are unchanged; their envy-free
comparisons and the protection of agents who stopped earlier remain
valid.

\begin{proposition}[One stage and its calls]
\label{prop:completion-stage}\label{lem:completion-counts}
A stage either finishes or leaves an envy-free allocation and a
nonempty proper group $\bar S$ such that every agent in $S=A\setminus\bar S$
dominates every member of $\bar S$. At active
size $k$, it uses at most $k^2$ snapshot calls, at most one shrinking call
with factor $2/\theta$, fewer than $k^4$ shrinking calls with factor
$2k^2/\tau$, and at most one successful final repair, possibly after
rejected attempts. The common shrinking call needs at most five
successful $s^8$-factor steps, and each separation call at most two.
A complete run has at
most one stage of each size and one base case. All other work,
including the final coverage and envy checks, uses recorded values.
\end{proposition}
\begin{proof}[Proof of \cref{thm:completion} and the proposition]
\emph{Correctness.} A stage finishes when its residue is empty.
Otherwise the global repair lets the agents in $S$ stop: they dominate
those who continue, and the current allocation is envy-free.
Induction on the active population, using
\cref{lem:base-cases}, proves complete envy-freeness.

\emph{Calls.} The snapshot and shrink counts follow from collection
and margin separation. Finishing with an empty residue needs no final
repair. Every other stage uses one returned repair. The step counts
follow from $s\ge2^{22}$ and $k\le s$:
\[
 2/\theta=6144s^{37}\le s^{38}\le(s^8)^5,
 \qquad 2k^2/\tau=8k^2s^{13}\le s^{16}=(s^8)^2.
\]

\emph{Termination and verification.} Every recursive call has fewer
agents. All requested snapshots, shrinks and repeated repairs terminate
almost surely after every preceding history, so the finite sequence
of stages does too. Recorded endpoint values suffice to check
disjointness, coverage up to cut endpoints and every envy inequality.
Only after these checks does the allocation become final.
\end{proof}

The call bounds are fixed before their outcomes are known. We may
therefore sum the procedures' conditional expected costs, with no
independence assumption between calls. \Cref{sec:accounting} combines
these bounds over the decreasing active populations.
\Needspace{10\baselineskip}
\section{Whole-bundle repair}\label{sec:kernel}

We now construct the procedure promised in \cref{thm:kernel}. It takes \emph{endowments}, bundles that may
be reassigned but must remain whole, together with additional cake. It
reassigns the endowments and distributes some of the cake to make the
allocation exactly envy-free. Later we prepare suitable endowments from
the cake itself, so the same repair also supplies partial allocations
from scratch.

Compensation can itself create envy: another agent may value a payment
more than its recipient does. We therefore choose the bundle assignment
and payment sizes together. We first work with numbers: we set aside a
\emph{reserve} of cake, measure values as fractions of each agent's value
of the reserve, and compute numerical payments, the \emph{credits}, that
remove all envy with room to spare. The room left by the credits must absorb the errors when we realise them
with reserve cake. These are the numerical and physical parts of the repair.
The empty-buffer case needs no compensation, as explained
in \cref{sec:overview}. For a nonempty buffer and $k\ge2$ agents, a trial
has five steps.

\begin{enumerate}[label=(\arabic*)]
\item \emph{Table and reserve.} Divide the available cake into pieces small
  for every agent. Draw a fixed number of independent sample points, refine at those
  points, and record all agents' values. Select disjoint batches of cells
  that preserve differences between agents' values; these supply the random
  additions. Independently reserve each original cell with
  the fixed reserve probability. Reject if a reserve value is outside its prescribed
  range, the reserve has too many refined cells, or a selected parent
  cell is reserved (\cref{sec:kernel-grid,sec:kernel-witnesses}).
  This step is the trial's \emph{preparation}.
\item \emph{Tentative bundles.} Create small random additions from the
  selected cells: for each agent pair in each batch, independently choose
  a uniform position in an ordered list of candidate additions. Draw an
  independent uniform permutation assigning the batches to endowments.
  Assign the enlarged bundles to maximise the recipients' total
  reserve-normalised value (\cref{sec:kernel-witnesses}).
\item \emph{Numerical compensation.} Compute credits that remove envy
  with room for payment errors. Reject if the compensation demands are
  inconsistent; otherwise the construction guarantees affordable
  credits and the required margins (\cref{sec:kernel-credits}).
\item \emph{Cake payments.} Prepare whole-cell payments, then fill each
  remaining deficit from the unused reserve, using a fresh uniform start
  in that recipient's value. Check the completed payment immediately:
  its owner receives her exact credit, and every observer's value stays
  within the allowed error. Reject on failure; otherwise remove the
  correction and continue (\cref{sec:kernel-lp,sec:kernel-crops}).
\item \emph{Verified output.} Perform the final checks of
  \cref{sec:kernel-certificate}, including envy-freeness and the whole
  assignment of every endowment. Any failed check
  rejects the trial without allocating cake.
\end{enumerate}
Both the envy margin and the allowed payment error scale with the
agents' disagreement on the reserve. Thus the query bound is unchanged
when valuations nearly agree.

Throughout this construction, $B$ is the repair buffer, the $W_j$ are
the endowments, and $d_i=v_i(B)$. The starting assignment satisfies
\eqref{eq:kernel-input}, with the fixed repair tolerance $\tau=1/(4s^{13})$.

\subsection{The reserve and the value table}\label{sec:kernel-grid}

The reserve must fund the payments while leaving cake for the tentative
bundles. We first record values on small pieces and select the pieces
needed for the random additions. Only then do we choose the reserve,
independently of the table and selected pieces. This order makes it clear that the
selection cannot favour pieces that will later be reserved.

The buffer $B$ contains an interval of positive length. Hungriness
therefore gives $d_i=v_i(B)>0$ for every participant: the same
construction applies to all $k$ agents.

To compare agents on a common scale, put $\nu_i=v_i/d_i$, so that
$\nu_i(B)=1$ and values express fractions of $B$. A common \emph{grid}
divides $B$ into \emph{cells} of normalised value at most the \emph{mesh}
$\eta$ for every agent; the mesh measures value, not length. To build it,
each agent $i$ cuts at multiples of $\eta d_i$ in her value along the
virtual order of $B$ (\cref{lem:union-cut}).
Combining these cuts gives the original, or \emph{coarse}, cells
$P_1,\ldots,P_{M_0}$ of
$\nu_i$-value at most $\eta$ for every $i$, with $1/\eta\le M_0\le2k/\eta$.

The numerical choices in \cref{def:kernel-parameters} use the fixed base
$s$ of \cref{thm:kernel}: the reserve probability is $q=s^{-11}$ and the
mesh is $\eta=\eta_E=s^{-23}$. The sample size is $m=s^{24}$.
Every call uses these same choices.

The table gives whole-cell values exactly. When a payment later takes only
part of a cell, its error will be controlled using random sample points
drawn now, before those later payment boundaries are chosen. The trial
draws $m$ independent \emph{sample points}
with distribution $\mu=k^{-1}\sum_{i=1}^k\nu_i|_B$, each by choosing an
agent uniformly and cutting at a uniform value target
(\cref{lem:sampling-average}). Since $\mu$ is the average of the
normalised valuations, it bounds each agent's value:
\begin{equation}\label{eq:kernel-mu-domination}
 v_i|_B\le kd_i\mu\qquad(i\in[k]).
\end{equation}
The trial then refines the cells at the sample points and records all
agents' endpoint values. The resulting record of every agent's
$v_i$-value of every refined cell is the \emph{value table}. A refined cell's
\emph{coarse parent} is the original cell containing it.

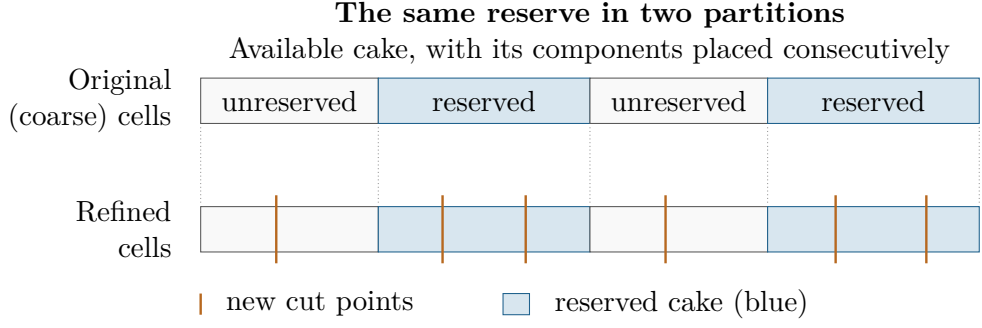
\begin{figure}[!htb]
\centering
\begin{tikzpicture}[x=1cm,y=1cm,font=\normalsize,>=Latex]
  \node[align=center] at (5.15,2.35)
    {\textbf{The same reserve in two partitions}\\
     Available cake, with its components placed consecutively};
  \node[anchor=east,align=right] at (-.25,1.45)
    {Original\\(coarse) cells};
  \node[anchor=east,align=right] at (-.25,-.25)
    {Refined\\cells};

  \foreach \yy in {1.15,-.55} {
    \draw[fill=gray!5,draw=black!65] (0,\yy) rectangle (2.35,\yy+.6);
    \draw[fill=aliceblue!18,draw=aliceblue!85!black]
      (2.35,\yy) rectangle (5.15,\yy+.6);
    \draw[fill=gray!5,draw=black!65] (5.15,\yy) rectangle (7.5,\yy+.6);
    \draw[fill=aliceblue!18,draw=aliceblue!85!black]
      (7.5,\yy) rectangle (10.3,\yy+.6);
  }
  \foreach \xx/\word in
    {1.175/unreserved,3.75/reserved,6.325/unreserved,8.9/reserved}
    \node at (\xx,1.45) {\word};

  \foreach \xx in {0,2.35,5.15,7.5,10.3}
    \draw[densely dotted,black!40] (\xx,1.10) -- (\xx,.10);
  \foreach \xx in {1.0,3.2,4.3,6.15,8.4,9.6}
    \draw[boborange!85!black,line width=.9pt] (\xx,-.70) -- (\xx,.20);

  \draw[boborange!85!black,line width=.9pt] (0,-1.38) -- (0,-1.10);
  \node[anchor=west] at (.2,-1.24) {new cut points};
  \draw[fill=aliceblue!18,draw=aliceblue!85!black]
    (4.0,-1.38) rectangle (4.35,-1.10);
  \node[anchor=west] at (4.55,-1.24) {reserved cake (blue)};
\end{tikzpicture}
\caption{A cell is the piece between consecutive cut points in the
available cake's order. The vertical orange ticks mark sample points that subdivide the original
cells. Both rows show the final reserve in blue; they are not successive
steps. The trial subdivides cells and selects representatives before
choosing which original cells to reserve.
Every refined part of a selected coarse cell stays in the reserve.
Components are drawn consecutively with gaps
omitted, so a cell may contain several intervals. Widths are schematic,
not values.}
\label{fig:cells-and-refinement}
\end{figure}

From this table, the trial selects disjoint batches of representative
pieces for the random additions. The construction in
\cref{sec:kernel-witnesses} uses only the input-normalised values $\nu_i$;
it requires no reserve values. The trial now reserves \emph{every}
coarse cell independently with probability $q$, whether selected or not.
The union $R$ of the reserved cells pays the credits. It performs the
\emph{mass check}: every reserve value must lie between half and three
halves of its mean. From then on it measures values relative to the
reserve:
\begin{equation}\label{eq:kernel-reserve-mass}
 qd_i/2\le r_i:=v_i(R)\le3qd_i/2\quad(i\in[k]),\qquad w_i:=v_i/r_i.
\end{equation}
After this normalisation $w_i(R)=1$, so a credit asks for the same
fraction of every agent's reserve value, not for equal original values.
\Cref{fig:cells-and-refinement} shows how the original cells and their
refinements describe the same reserve.
\FloatBarrier

The trial next checks that the refined reserve has at most $N_*:=1024qm=1024s^{13}$
cells $R_1,\ldots,R_{N_R}$, each of $w_i$-value at most $\eta_R:=2\eta/q$.
It also rejects if any selected representative piece is reserved. These
checks do not alter the reserve choices or the earlier selection.
From the table it computes two measures of the disagreement between agents
$i$ and $j$ on the reserve, the largest difference on a cell and the total
absolute difference:
\begin{equation}\label{eq:kernel-disagreement}
 A_{ij}=\max_\ell|(w_i-w_j)(R_\ell)|,\qquad
 D^R_{ij}=\sum_\ell|(w_i-w_j)(R_\ell)|\le N_*A_{ij}.
\end{equation}
The largest difference sets the required assignment advantage; the sum
bounds errors when payments are cut. The next lemma verifies the reserve
checks and records the independence used in the analysis.

\begin{lemma}[Reserve and table]\label{lem:kernel-grid}\label{lem:app-kernel-reserve}
The refined table of $v_i$-values is independent of reserve labels. The
mass check fails with probability at most $4k\eta/q$; the cell-count
check fails with probability at most $1/512$. After the mass check
passes, every refined cell has
$w_i$-value at most $\eta_R$.
\end{lemma}
\begin{proof}
\emph{Independence.} The grid and table are fixed before the independent
reserve labels are drawn. Conditional on the grid, the sample points are
independent with distribution $\mu$ by \cref{lem:sampling-average}.
For the analysis we may therefore fix the reserve labels first and still
use that sample distribution. This does not allow us to condition on
the later checks, which depend on both the table and the labels.

\emph{The mass check.} For each agent $i$, the value $\nu_i(R)$
has mean $q$, its contributions from individual cells are independent and
at most $\eta$, and so its variance is at most $q\eta$. Chebyshev's
inequality bounds the probability of a deviation greater than $q/2$ by
$4\eta/q$, and adding this bound over agents gives $4k\eta/q$.

\emph{The number of refined reserve cells.} Fix the entire refined
value table, before the reserve labels. It has at most $M_0+m\le2m$
cells. Each belongs to a parent reserved with probability $q$, so
linearity of expectation gives $\E[N_R\mid\text{table}]\le2qm$.
Cells with the same parent share a label; no independence between them
is needed. Markov's inequality now gives
\[
 \Pr[N_R>1024qm\mid\text{table}]\le\frac1{512}.
\]
Averaging over the table proves the required bound, including the event
that the mass check also passes.

\emph{After the mass check.} Every refined cell $E$ lies in a coarse cell,
so $\nu_i(E)\le\eta$; since $\nu_i(R)\ge q/2$ after the mass check,
$w_i(E)\le\eta/\nu_i(R)\le\eta_R$.
\end{proof}

\subsection{Separation of assignment values}\label{sec:kernel-witnesses}
\label{sec:kernel-isolation}

The tentative bundles must leave room for compensation. Maximising the
recipients' total reserve-normalised value rules out any exchange that
raises the sum, but an exchange may still lose almost nothing. We need a
definite loss: adding compensation requirements around a cycle cancels
each credit, so the loss of the cyclic exchange must pay for the required
margins. It must scale with reserve disagreement, which also governs
payment errors.

We create this separation by adding small random pieces to the endowments
before assigning them. A piece valued differently by two agents changes
the comparison between giving its bundle to one of them or to the other.
Arbitrary pieces could miss the few cells on which agents disagree, so we
first select cells that retain these disagreements. For each pair, we
then arrange possible additions in a list whose preference difference
increases by a definite amount at each step. Choosing a random place in
that list makes a near tie unlikely. We now describe the selection made
before the reserve labels, followed by the random additions made after
the reserve checks.

\smallskip\noindent\textbf{Representative cells.}
For each pair of agents, we select at most two candidate cells at a time,
called \emph{representatives}, before the reserve values are known. After
normalisation, one of them will retain at least one third of the largest
reserve-cell difference $A_{ij}$; this cell is a \emph{witness} for the pair.
Repeated selections supply distinct witnesses. Adding a positive-difference
witness, or removing a negative-difference one, increases the assignment
comparison. The representatives must be selected without using reserve
labels, since selected cake cannot also be used for payments.

Two cells suffice for one pair. First choose a cell that the first agent
values most. Then choose a cell with the largest absolute departure from
the ratio seen on that first cell, where the departure is the second
agent's value minus the first agent's value times that ratio. If the
departure is zero everywhere, the first cell alone suffices. The next
elementary lemma shows that these choices still work after the two
valuations are divided by their different reserve values, which the
selection does not consult. In the lemma, $x_E,y_E$ are the two recorded
values of cell $E$, and the coefficients $u,v$ allow their later
rescaling. \Cref{fig:pair-representatives} illustrates the two
choices.

\begin{figure}[!htb]
\centering
\textbf{A worked example of the two choices}\par\medskip
\begingroup
\renewcommand{\arraystretch}{1.35}
\setlength{\tabcolsep}{10pt}
\begin{tabular}{@{}m{6cm}*{3}{>{\centering\arraybackslash}m{1.6cm}}@{}}
\toprule
& \colorbox{aliceblue!12}{\strut Cell $F$}
& \colorbox{boborange!14}{\strut Cell $G$} & Cell $E$ \\
\midrule
Alice's value & $0.50$ & $0.30$ & $0.20$ \\
Bob's value & $0.10$ & $0.60$ & $0.30$ \\
\addlinespace[3pt]
Bob's value minus\newline one fifth of Alice's value
  & $0$ & $0.54$ & $0.26$ \\
\bottomrule
\end{tabular}
\endgroup
\par\medskip
\parbox{.94\linewidth}{\centering
First choose $F$: Alice values it most, and Bob's value there is
one fifth of hers. Then choose $G$: it has the largest absolute
 departure from that ratio.}
\caption{Two representative cells, with illustrative values rather than
a protocol run. Blue marks the first choice, $F$; orange marks the
second, $G$. These calculations select cells; they do not cut or mix
them. \Cref{lem:representative-basis} proves why the choice remains useful after the
agents' values are divided by different reserve totals: at least one
selected cell retains one third of the largest absolute difference
on any candidate cell.}
\label{fig:pair-representatives}
\end{figure}

\Needspace{9\baselineskip}
\begin{lemma}[Two representative cells]\label{lem:representative-basis}
A nonempty finite list of nonnegative pairs $(x_E,y_E)$ has a sublist
$T$ of at most two pairs such that, for every $u,v\in\mathbb R$,
\[
 |u x_E-v y_E|\le3\max_{F\in T}|u x_F-v y_F|
 \qquad\text{for every listed }E.
\]
The sublist is found by two scans and scalar arithmetic, with a fixed
rule for ties.
\end{lemma}
\begin{proof}
Choose $F$ maximising $x_F$. If $x_F=0$, a pair maximising $y_E$
satisfies the bound immediately. Otherwise put $r_E=y_E-(y_F/x_F)x_E$, the departure from the ratio on $F$,
and choose $G$ maximising
$|r_G|$. Retain only $F$ if $r_G=0$, and $F,G$ otherwise.
For $T_E=ux_E-vy_E$,
\[
 T_E=(x_E/x_F)T_F-vr_E,\qquad
 |vr_E|\le|vr_G|=|(x_G/x_F)T_F-T_G|.
\]
Both ratios lie in $[0,1]$, so
$|T_E|\le2|T_F|+|T_G|\le3\max\{|T_F|,|T_G|\}$.
When $r_G=0$, the first identity directly gives $|T_E|\le|T_F|$,
justifying the singleton choice.
\end{proof}

The trial applies this lemma to the recorded pairs $(\nu_i(E),\nu_j(E))$,
without consulting the reserve values. We shall give each endowment its
own batch of representative cells. We use independent random choices
within these disjoint batches, then a random permutation to treat the
endowments symmetrically.

For one batch, visit the unordered agent pairs in a fixed
order. For each pair, make $L=s^8$ selections. After each selection,
immediately remove the coarse parents of its at most two representatives
from the available collection. Thus different selections, including
selections for different pairs, use disjoint parents. A batch contains
at most $k(k-1)L\le k^2L$ selected parents. Repeat this construction
$k$ times on what remains, giving disjoint batches of cake available for
the random additions.

After the reserve checks described in \cref{sec:kernel-grid}, if no selected parent is
reserved, every reserve cell was available at every selection. For a
pair $i<j$, choose from each of its selected sets a cell maximising
$|(w_i-w_j)(E)|$, with fixed ties. Denote these $L$ witnesses by
$E_1,\ldots,E_L$. This last choice uses the reserve normalisation but no
new queries. Unchosen representatives stay unused. The next lemma
explains why these witnesses retain the disagreement we need.

\Needspace{8\baselineskip}
\begin{lemma}[Witnesses outside the reserve]\label{lem:kernel-witness}
Every selection has positive value available for every agent.
The probability that any selected parent is reserved is at most
$qk^3L$. If none is reserved and the mass check passes, the chosen
witnesses for every distinct pair $i,j$ in each batch satisfy
$|(w_i-w_j)(E_r)|\ge A_{ij}/3$ for $r=1,\ldots,L$.
These witnesses have distinct coarse parents.
After the mass check, each batch has $w_i$-value at most
$k^2L\eta_R\le1/(4s^2)$ for every agent $i$.
\end{lemma}
\begin{proof}
At most $k^3L$ parents are removed, of total $\nu_i$-value at most
$k^3L\eta<1/2$ by \cref{lem:app-kernel-closure}. Every agent
therefore still values the remaining collection positively. The selected
parents depend only on the table, which is independent of reserve labels
(\cref{lem:kernel-grid}). Conditional on that table, each parent is
reserved with probability $q$, giving the union bound $qk^3L$.

If no selected parent is reserved, a reserve cell attaining $A_{ij}$
remains available at every selection. Applying \cref{lem:representative-basis}
with $u=1/\nu_i(R)$ and $v=1/\nu_j(R)$ shows that one selected cell has
difference at least $A_{ij}/3$. Removing its whole parent makes the
witnesses distinct. Finally, one batch has at most $k^2L$ parents, each
of $w_i$-value at most $\eta_R$. The last bound follows from
\cref{lem:app-kernel-closure}.
\end{proof}

\smallskip\noindent\textbf{Random additions.}
Consider the witnesses for one pair $i<j$ in one batch, and write
$a_r=(w_i-w_j)(E_r)$. We want to make the difference between the pair's
values increase in controlled steps. Start with all witnesses having
negative difference included. At step $r$, add $E_r$ if $a_r>0$ and
remove it if $a_r<0$; omit a zero-difference witness throughout. Each
step increases the difference by $|a_r|$.

Independently for each pair in each batch, choose a random position
(or \emph{threshold}) $T$ uniformly from $\{0,\ldots,L\}$ and use the piece after step $T$:
\[
 \bigcup_{r\le T,\ a_r>0}E_r
 \ \cup\!
 \bigcup_{r>T,\ a_r<0}E_r.
\]
This is one allowed uniform-real draw followed by comparisons with the
$L+1$ equal subintervals of $[0,1]$; a fixed boundary convention suffices.
The selected parents are disjoint, so choices for different pairs never
try to use the same cake. Unite their chosen witnesses to obtain the
piece from that batch.

For illustration, take three witness differences $2,-3,1$ in common units. They
give the following list.
The values increase even though the physical pieces are not nested:
\[
\begin{array}{c|cccc}
 T&0&1&2&3\\ \hline
 \text{included witnesses}&E_2&E_1\cup E_2&E_1&E_1\cup E_3\\
 \text{preference difference}&-3&-1&2&3
\end{array}
\]
A short interval can contain only a few entries of a list whose values
are separated in this way. That elementary observation will replace a
probability estimate over all possible assignments.

Finally choose an independent uniform permutation of the $k$ batch
pieces, and attach them to $W_1,\ldots,W_k$ in that order. Write $Z_j$
for the piece attached to $W_j$, called its \emph{seed piece}. The
thresholds separate assignment values; the permutation makes the
additions symmetric among endowments. Each seed piece is small because
its whole batch is small, without any concentration estimate.

\begin{lemma}[Symmetry of the additions]\label{lem:kernel-symmetry}
Conditional on all data before the thresholds and permutation,
$(Z_1,\ldots,Z_k)$ and $(Z_{\pi(1)},\ldots,Z_{\pi(k)})$ have the same
distribution for every permutation $\pi$ of $[k]$.
\end{lemma}
\begin{proof}
Even after the batch pieces have been chosen, the independent uniform
permutation assigns them symmetrically to the endowments.
\end{proof}

\smallskip\noindent\textbf{Assignment gaps.}
In the assignment matrix $V$, row $i$ represents agent $i$ and column $j$
represents $W_j\cup Z_j$. Set $V_{ij}=w_i(W_j\cup Z_j)$.
Using a fixed rule for ties, the trial chooses a permutation
$\pi$ maximising $\sum_iV_{i,\pi(i)}$, and calls $G_i=W_{\pi(i)}\cup Z_{\pi(i)}$
the \emph{tentative bundle} of agent $i$.

Consider passing bundles around a cycle of agents, and call the decrease in
their total reserve-normalised value the \emph{loss} of the cycle. We want
this loss to cover the margins required for compensation, even though
individual agents may still prefer another bundle. We ask for a final
margin $\lambda A_{ij}$, where $\lambda=s^4/8$ is the multiplier fixed in
\cref{def:kernel-parameters}. The credits will provide twice this margin,
so the error budget is simple:
\begin{equation}\label{eq:kernel-scale}
 \underbrace{2\lambda A_{ij}}_{\text{advantage with credits}}
 -\underbrace{\lambda A_{ij}}_{\text{total payment error}}
 =\underbrace{\lambda A_{ij}}_{\text{remaining advantage}}.
\end{equation}
For the probability estimate, we give half the error budget to whole
cells and half to their corrections. The actual check will concern
their combined payment. The multiplier $\lambda$ is chosen large enough to
leave room for these errors while keeping the margin proportional to
disagreement. No positive margin is required when $A_{ij}=0$.

There may be many cyclic exchanges, and their scores share the same
random pieces. Instead, fix a bundle and compare the best total score
when it must go to one recipient with the best score when it must go to
another. If a cyclic exchange loses very little, one of these comparisons
must be close. There are at most $k^3$ comparisons to consider.

This follows the idea behind the \emph{isolation method}
\cite{MulmuleyVaziraniVazirani1987}: random changes to individual
contributions separate competing optimal solutions, even though the scores
of complete solutions are dependent. Here the changes are actual cake
additions, and we need a gap proportional to the reserve disagreement.

Let $\Phi_{ij}$ be the best assignment score after row $i$ and column $j$
are deleted. Then $V_{ij}+\Phi_{ij}$ is the best score among assignments
that give bundle $j$ to agent $i$. These constrained optima are used only
in the proof; the protocol need not compute them. The next lemma says
that different possible recipients give well-separated scores whenever
they disagree on a reserve cell.

\begin{lemma}[Separation of assignment scores]\label{lem:kernel-isolation}\label{lem:app-kernel-forced}\label{lem:app-kernel-smallball}
Outside an event of probability at most $k^3(12k\lambda+1)/(L+1)$,
every trial reaching
the assignment stage satisfies
\begin{equation}\label{eq:kernel-assignment-gaps}
 |V_{ij}+\Phi_{ij}-V_{\ell j}-\Phi_{\ell j}|>2k\lambda A_{i\ell}
\end{equation}
for every pair $i,\ell$ with $A_{i\ell}>0$ and every bundle $j$.
\end{lemma}
\begin{proof}
Call the expression inside the absolute value the \emph{tested difference}.
A comparison is \emph{close} when its absolute value is at most
$2k\lambda A_{i\ell}$. We bound each such probability and then add the bounds.

Fix the table, reserve labels and permutation assigning batches to
endowments, and suppose preparation passes. For a pair $i,\ell$
and column $j$, fix every threshold except the one for that pair in the
batch assigned to $j$. Only column $j$ can now change. Both scores
$\Phi_{ij}$ and $\Phi_{\ell j}$ omit that column, so they stay fixed
even though its entries for other agents may change. The tested difference
therefore varies only through the chosen pair's value difference in column $j$.

Orient the tested pair as in its witness selection; reversing its order
only reverses the sign of the tested difference. The remaining random
part is a constant plus
\[
 S(t)=\sum_{r\le t,\ a_r>0}a_r+\sum_{r>t,\ a_r<0}a_r,
 \qquad t=0,\ldots,L.
\]
By \cref{lem:kernel-witness}, successive values differ by
$|a_t|\ge A_{i\ell}/3$. An interval of width $4k\lambda A_{i\ell}$ therefore
contains at most $\lfloor12k\lambda\rfloor+1$ of these values. Since the
threshold is uniform, the comparison is close with probability at most
\[
 \frac{12k\lambda+1}{L+1}.
\]
This bound holds for every fixed choice of the other thresholds and the
permutation. Average over those choices and add over at most $k^3$
comparisons, giving the stated bound. When $A_{i\ell}=0$, the lemma requests no
gap and this division is not used.

\end{proof}

\subsection{Numerical credits and envy margins}\label{sec:kernel-credits}

We first calculate compensation with numbers. A \emph{credit} $t_i$
adds the same number to every agent's reserve-normalised value of $G_i$;
later we realise this ideal payment with cake, allowing a controlled error
for other agents. The credit must cover agent $i$'s
preference for another bundle, that recipient's credit, and the error
margin.

These requirements could chain around a cycle of agents and force the
credits ever higher. When they are added around a cycle, however, the
credits cancel, leaving the exchange gain plus the requested margins, and
the assignment gaps of \cref{lem:kernel-isolation} will make this sum
nonpositive. Each agent can then receive the largest cumulative demand
along a chain ending at her. Formally, for $i,l\in[k]$, give the arrow $l\to i$ the cost
$\psi_{il}=w_i(G_l)-w_i(G_i)+2\lambda A_{il}$, which is the required excess of
the destination's credit over the source's. Reject a positive-cost cycle.
Otherwise let the credit $t_i$ be the maximum of $0$ and the costs of
simple paths (paths visiting no agent twice) ending at $i$. These are the longest-path subsidies of
Halpern and Shah~\cite[Theorem~2]{HalpernShah2019}, with arrows reversed
so that paths end at their recipient, and with margins added to the
costs. A credit may be zero: the repair needs to preserve
the endowments and remove envy, not give every agent a positive payment.

Affordability has a deterministic explanation. Before reassignment, each
agent's own bundle is close to her favourite, even after the small random
additions. Maximising the total value cannot make the sum of these
shortfalls larger. A compensation path visits each agent at most once,
so its demands are bounded by that total shortfall, together with the
small margins. This uses small initial envy, not near equality of all
endowment values.

If the graph has no positive-cost cycle, its path credits satisfy
\begin{equation}\label{eq:kernel-credit-checks}
 0\le t_i\le\frac{k}{s^2},
\end{equation}
\begin{equation}\label{eq:kernel-money}
 V_{i,\pi(i)}+t_i-V_{i,\pi(j)}-t_j\ge2\lambda A_{ij}
 \qquad(i,j\in[k]).
\end{equation}
Thus this stage rejects only for a positive-cost cycle and needs no new
queries. The following lemma proves affordability and the margins, then
uses assignment separation to rule out rejection; it extends the subsidy
principle of \cite[Theorem~1]{HalpernShah2019}.

\begin{lemma}[Affordable credits]\label{lem:kernel-credits}\label{lem:app-kernel-bands}
Suppose preparation passes. If the credit graph has no positive-cost
cycle, its path credits satisfy \eqref{eq:kernel-credit-checks} and
\eqref{eq:kernel-money}. In particular, $t_i<1/(16k)$ and
$\sum_i t_i<1/16$. The assignment gaps
\eqref{eq:kernel-assignment-gaps} guarantee that no positive-cost cycle
exists.
\end{lemma}
\begin{proof}
\emph{Credits meet the margins.} Suppose there is no positive-cost
cycle. Extend a maximum-cost path ending at $\ell$ by the arrow
$\ell\to i$; use the length-zero path when $t_\ell=0$. Deleting any
resulting nonpositive cycle cannot lower its cost. Thus
$t_i\ge t_\ell+\psi_{i\ell}$,
which is exactly \eqref{eq:kernel-money}.

\emph{Credits are affordable.} The input condition and mass check give
$w_i(W_j)-w_i(W_i)\le2\tau/q=1/(2s^2)$. Every seed piece is worth
between zero and $1/(4s^2)$, so $V_{ij}-V_{ii}\le3/(4s^2)$.
Write $M_i=\max_jV_{ij}$ for agent $i$'s favourite-bundle value.
Her shortfall is $M_i-V_{i,\pi(i)}$. Optimal assignment cannot increase
the total shortfall, because the identity assignment was available:
\[
 \sum_h(M_h-V_{h,\pi(h)})\le\sum_h(M_h-V_{hh})
 \le\frac{3k}{4s^2}.
\]
An arrow entering $i$ costs at most $M_i-V_{i,\pi(i)}+2\lambda\eta_R$.
A simple compensation path charges each agent at most once and has at
most $k-1$ arrows. Since $2\lambda\eta_R\le1/(4s^2)$ by
\cref{lem:app-kernel-closure},
\[
 t_i\le\sum_h(M_h-V_{h,\pi(h)})+(k-1)2\lambda\eta_R
 \le\frac{3k}{4s^2}+\frac{k-1}{4s^2}<\frac{k}{s^2}.
\]
This proves \eqref{eq:kernel-credit-checks};
\cref{lem:app-kernel-closure} gives the stated individual and total
bounds. No additional probability estimate is needed.

\emph{Assignment gaps prevent rejection.} Now suppose
\eqref{eq:kernel-assignment-gaps} holds. Consider a simple cycle of arrows
in the credit graph, and let $\Delta$ be the loss in total assignment
score when each recipient takes the bundle at the source of her incoming
arrow. For an arrow $\ell\to i$, the current assignment gives column
$j=\pi(\ell)$ to $\ell$ and has optimal score $\mathrm{OPT}$. The cyclic
reassignment gives that column to $i$, so
\[
 \mathrm{OPT}-\Delta\le V_{ij}+\Phi_{ij}\le\mathrm{OPT}
 =V_{\ell j}+\Phi_{\ell j}.
\]
If some $A_{i\ell}$ on the cycle is positive, apply
\eqref{eq:kernel-assignment-gaps} to an arrow with the largest such
value. It gives $\Delta>2k\lambda\max A_{i\ell}\ge2\lambda\sum A_{i\ell}$,
since the cycle has at most $k$ arrows. If all these values are zero,
assignment optimality gives $\Delta\ge0$. The ordinary envy terms sum
to $-\Delta$: they compare exactly the cyclic reassignment with the chosen
assignment. Thus the total arrow cost, $2\lambda\sum A_{i\ell}-\Delta$,
is nonpositive in both cases.

\end{proof}

\Needspace{8\baselineskip}
\subsection{Payments with reserve cake}\label{sec:kernel-lp}

A credit is a number; we must pay it with disjoint cake. Giving each
recipient her credited fraction of every reserve cell would have the
right value for every agent. But a numerical fraction does not specify
an actual subpiece: cutting that fraction of a cell's length, or of one
agent's value, need not preserve the same fraction for another agent.

We first pay with whole cells, leaving only a small deficit for each
recipient. We then fill those deficits, one at a time, from the unused
reserve. Each correction is checked immediately. The check ensures
that the \emph{completed payment} gives its recipient exactly her credit,
and that every other agent's value is close to that credit. This is the
guarantee the final envy calculation needs. It also leaves enough reserve
for the next recipient.

For each reserve cell $R_\ell$, let $x_{\ell j}$ be the fraction intended
for recipient $j$, and let $x_{\ell0}$ be the unused fraction.
The numerical plan requires nonnegative fractions with
$\sum_{j=0}^k x_{\ell j}=1$ and
\begin{equation}\label{eq:kernel-lp}
 \sum_\ell w_i(R_\ell)x_{\ell j}=t_j
 \qquad(i,j\in[k]).
\end{equation}
The constant fractions $x_{\ell j}=t_j$, leaving the rest unused, meet
these requirements because $w_i(R)=1$ and $\sum_jt_j<1/16$.

A cell is \emph{split} if more than one fraction is positive; no physical
cut has yet been made. There are at most $k^2$ equations besides the
cell totals. The elementary elimination argument of
\cref{lem:sparse-fractions} preserves those equations while leaving
at most $k^2$ split cells. Give recipient $j$ all cells assigned wholly
to her, calling their union $Y_j^{\rm whole}$. Leave the split cells
unused. Her remaining deficit is
$\epsilon_j=t_j-w_j(Y_j^{\rm whole})$. The next lemma bounds both the
deficits and the disagreement introduced by discarding split cells.

\begin{lemma}[Whole-cell payments]\label{lem:kernel-lp}
At most $k^2$ cells remain split. For all $i,j\in[k]$,
\[
 0\le w_i(Y_j^{\rm whole})\le t_j,\qquad
 \epsilon_j\ge0,\qquad \sum_j\epsilon_j\le k^2\eta_R,
\]
\[
 |(w_i-w_j)(Y_j^{\rm whole})|\le k^2A_{ij}\le \lambda A_{ij}/2.
\]
\end{lemma}
\begin{proof}
Let $I_{\rm split}$ be the discarded split cells. The residual identity
\eqref{eq:rounding-residual} is
\[
 w_i(Y_j^{\rm whole})=t_j-
 \sum_{\ell\in I_{\rm split}}x_{\ell j}w_i(R_\ell).
\]
The removed terms are nonnegative. Since each cell is worth at most
$\eta_R$ and its recipient fractions sum to at most one,
\[
 \sum_j\epsilon_j\le
 \sum_{\ell\in I_{\rm split}}\eta_R\sum_jx_{\ell j}
 \le k^2\eta_R.
\]
There is no extra factor for recipients. Subtracting the identities for
$i$ and $j$ cancels $t_j$ and bounds disagreement by $k^2A_{ij}$;
$k^2\le\lambda/2$ by \cref{lem:app-kernel-closure}.
\end{proof}

\paragraph{Exact correction pieces.}\label{sec:kernel-crops}
\label{sec:app-kernel-correction}
Set $S=R\setminus\bigcup_jY_j^{\rm whole}$, the currently unused reserve.
Process recipients in a fixed order. For a positive deficit
$\epsilon_j$, place the intervals of $S$ end to end and regard their
order as circular. Choose a fresh uniform start in recipient $j$'s
value, and take the next $\epsilon_j$ units of $w_j$-value, wrapping if
necessary. Denote the resulting piece by $C_j$. This is the circular
piece of \cref{lem:sampling-crop}; it follows the intervals of $S$ in
order, skipping the gaps between them. A zero deficit uses
$C_j=\emptyset$.

Record the new endpoints for every original agent, and immediately check
the completed payment $P_j=Y_j^{\rm whole}\cup C_j$:
\begin{equation}\label{eq:kernel-correction-check}
 w_j(P_j)=t_j,\qquad
 |w_i(P_j)-t_j|\le\lambda A_{ij}\quad(i\in[k]).
\end{equation}
On failure, reject the whole trial. Otherwise remove $C_j$ from $S$ and
continue. The following argument shows that every reached correction
has ample cake available; its probability analysis then shows that the
checks usually pass.

\begin{lemma}[Completed payments]\label{lem:kernel-crops}\label{lem:app-kernel-crops}
Before every reached correction, and after every accepted correction,
more than $7/8$ of every agent's reserve value remains unused.
Every deficit therefore fits. If all payments pass, they are disjoint,
satisfy \eqref{eq:kernel-correction-check}, and obey
\[
 \sum_jw_i(P_j)<\frac18\qquad(i\in[k]).
\]
The probability that a reached payment check fails is at most
\[
 \frac{8k^4(1+b)}{qb(m+1)}+\frac{2k^3\eta_R}{b},
 \qquad b:=\frac{\lambda}{4N_*}.
\]
This probability averages over the table and all the trial's other
choices; it does not condition on passing the preparation checks.
\end{lemma}
\begin{proof}
\emph{Enough cake for each correction.} Let $J$ contain the recipients
whose completed payments have passed. The used reserve consists of
their payments and the whole-cell payments of recipients still waiting.
For any observer $i$, the completed-payment check and
\cref{lem:kernel-lp} give
\[
 \begin{aligned}
 w_i(\text{used reserve})
 &=\sum_{j\in J}w_i(P_j)+\sum_{j\notin J}w_i(Y_j^{\rm whole})\\
 &\le\sum_jt_j+\lambda\sum_{j\in J}A_{ij}
 <\frac1{16}+k\lambda\eta_R<\frac18.
 \end{aligned}
\]
Here $A_{ij}\le\eta_R$, and \cref{lem:app-kernel-closure} gives the
last inequality. This proves the supply bound before the next correction
and after any accepted one, as well as the final total-payment bound.
Since $\epsilon_j\le t_j<1/16$, every deficit fits. A failed check
rejects the trial; no supply claim is needed after that rejection.

\emph{One fixed cell estimate throughout the corrections.} Write
$\gamma_{ij}=(w_i-w_j)|_R$ for the actual signed disagreement. The table
specifies it exactly on whole cells, but a correction may use only part
of a cell. As in \cref{def:sampling-model},
join its recorded cumulative values by straight lines inside the
original sampled cells, giving the cell estimate $\widehat\gamma_{ij}$ with
\[
 |\widehat\gamma_{ij}|(R)=D^R_{ij}.
\]
Only the estimate is linear inside cells; actual valuations need not be.
We control two quantities: the difference between actual and estimated
disagreement on a correction, and the estimated disagreement itself.
The sampled grid controls the first through the correction's boundaries;
the fresh random start controls the second. Keep the estimate fixed as
corrections remove pieces: their new boundaries do not define a new
interpolation grid.

\emph{Estimated disagreement in a correction.} Fix the full history
before recipient $j$'s fresh start. The circular inclusion identity
\eqref{eq:sampling-model-crop}, $w_j(S)>7/8$ and $S\subseteq R$ give
\[
 \E\!\left[|\widehat\gamma_{ij}(C_j)|
       \,\middle|\,\text{information before this start}\right]
 \le\frac{\epsilon_j}{w_j(S)}|\widehat\gamma_{ij}|(S)
 \le2\epsilon_jD^R_{ij}.
\]
The circular inclusion identity applies to the estimate as well as to
the agents' values; it does not require them to agree within cells.
If $D^R_{ij}=0$, the estimate vanishes.
Otherwise Markov's inequality bounds the chance of exceeding $bD^R_{ij}$
by $2\epsilon_j/b$. The deficits were fixed before corrections started,
and their sum is at most $k^2\eta_R$. Sum the conditional bounds over
reached recipients and observers, then average over preparation. The
resulting event $\mathcal M$ satisfies
\[
 \Pr[\mathcal M]\le\frac{2k^3\eta_R}{b}.
\]
Stopping early can only reduce this sum.

\emph{Errors at the endpoints.} For analysis, first fix the coarse
reserve labels, independently of the sample points. If the mass check
fails, there are no corrections. Otherwise the signed measures
$\gamma_{ij}$ for $i<j$ are fixed before sampling. There are
at most $k^2/2$ of them; reversing the agents reverses the sign. A later
preparation rejection gives an empty endpoint list.

At a reached correction, condition on the full history before its fresh
start. Each of its two new boundaries has marginal distribution
$w_j|_S/w_j(S)$; the boundaries need not be independent. The deterministic
supply bound just proved and \eqref{eq:kernel-mu-domination} give
\[
 \frac{w_j|_S}{w_j(S)}\le2w_j\le\frac{4k}{q}\mu.
\]
Each correction supplies one group of at most two boundaries, generated
by its fresh start. Thus there are at most $H=2k$ boundaries, with
conditional marginal laws bounded by $\Lambda\mu$, where
$\Lambda=4k/q$. These bounds hold after every preceding history,
including the sampled grid and stopping after a failed check. Count the
boundaries of a rejecting correction as well. Inherited boundaries are
already counted; original grid boundaries have zero interpolation error.

Apply \cref{prop:sampling-endpoints} with these $H,\Lambda$ and the stated $m,b$.
Averaging its conclusion over the independent reserve
labels gives a joint failure event $\mathcal B$ with
\[
 \Pr[\mathcal B]\le\frac{8k^4(1+b)}{qb(m+1)}.
\]
Outside it, \cref{cor:sampling-unions} gives, for every reached correction,
\[
 |\gamma_{ij}(C_j)-\widehat\gamma_{ij}(C_j)|\le bD^R_{ij}.
\]
This covers pieces with inherited boundaries as well as new ones. No
favourable sampled table has been conditioned upon.

Neither probability estimate conditions on the other succeeding.
Outside their union,
\[
 |\gamma_{ij}(C_j)|\le2bD^R_{ij}
 \le2bN_*A_{ij}=\lambda A_{ij}/2.
\]
This includes zero recorded disagreement, where it gives exact equality.
Adding the whole-cell error from \cref{lem:kernel-lp} gives
$|w_i(P_j)-t_j|\le\lambda A_{ij}$. The owner-value identity holds
for every start by \cref{lem:sampling-crop}. Thus every reached
completed-payment check passes
outside the two failure events. This argument concerns the process
with these checks: its accepted payments establish the supply bound
used for each later random start.
\end{proof}

\subsection{Output guarantees and query cost}\label{sec:kernel-certificate}
\label{sec:app-kernel-execution}

Set $Y_i=Z_{\pi(i)}\cup P_i$ for every agent. The output is
$X_i=W_{\pi(i)}\cup Y_i$. The completed-payment check gives exactly
the representation needed to transfer the numerical envy margin to
the actual allocation. For $i,j\in[k]$, put
$e_{ij}=w_i(P_j)-t_j$. Then
\begin{equation}\label{eq:kernel-representation}
\begin{gathered}
 w_i(X_j)=V_{i,\pi(j)}+t_j+e_{ij},\qquad e_{jj}=0,\\
 e_{ij}=(w_i-w_j)(P_j),\qquad
 |e_{ij}|\le \lambda A_{ij}.
\end{gathered}
\end{equation}
Using recorded values, the trial then checks
\eqref{eq:kernel-representation}, the reserve and credit bounds, that each
endowment is assigned exactly once, that the pieces are disjoint, that the
allocation is exactly envy-free, and that
$v_i(\bigcup_jY_j)\le v_i(B)/4$ for every participant. It returns only
if every check passes.

\Needspace{12\baselineskip}
\begin{proposition}[Exact output and remaining margin]\label{prop:kernel-output}
For a nonempty buffer, every returned trial assigns the endowments whole
and returns an exactly
envy-free allocation $X_i=W_{\pi(i)}\cup Y_i$, with disjoint finite-union
additions $Y_i\subseteq B$. In particular,
$v_i(X_i)\ge\max_jv_i(W_j)$ for every agent.
For every participant $i$,
\[
 v_i\Bigl(\bigcup_jY_j\Bigr)\le\tfrac14v_i(B).
\]

The reserve and payment data satisfy
\eqref{eq:kernel-reserve-mass}, \eqref{eq:kernel-scale},
\eqref{eq:kernel-money} and \eqref{eq:kernel-representation}. In particular,
\[
 v_i(X_i)-v_i(X_j)\ge \lambda r_iA_{ij}
 \qquad(i,j\in[k]).
\]
Here $Z_j\subseteq B\setminus R$, $t_i\ge0$ and $A_{ij}=A_{ji}\ge0$;
the reserve has at most $N_*$ refined cells, each of $w_i$-value at most
$\eta_R$.
\end{proposition}
\begin{proof}
The credit margin and payment error give
\[
 w_i(X_i)-w_i(X_j)\ge2\lambda A_{ij}-e_{ij}
 \ge\lambda A_{ij}\ge0.
\]
Multiplying by $r_i$ gives the stated margin. Seeds lie outside the reserve;
whole-cell payments are disjoint, and every correction is removed before
the next is cut. Each endowment remains whole in some $X_l$, so
nonnegative valuations and envy-freeness give
$v_i(X_i)\ge v_i(X_l)\ge v_i(W_j)$ for every $j$.
All additions lie in the reserve or in at most $k^3L$ selected parents.
Consequently,
\[
 v_i\Bigl(\bigcup_jY_j\Bigr)
 \le\left(k^3L\eta+\frac{3q}{2}\right)v_i(B)
 <\frac14v_i(B),
\]
by \cref{lem:app-kernel-closure}. The remaining listed properties are
among the final checks.
\end{proof}

The small-total-addition guarantee is for participating agents. For
any original nonparticipant, the general fact needed later is simply
$v_o(\bigcup_jY_j)\le v_o(B)$: all added cake comes from the buffer.

\label{rem:kernel-dipole}

\paragraph{Success probability and query cost.}
The final checks guarantee correctness whenever a trial returns. For
the expected query bound, we must also show that it usually returns. We
combine the five rejection bounds below; the resulting bound applies
after any earlier history, permitting retries and calls from other procedures.
\begin{lemma}[Bounds for a general trial]\label{lem:kernel-trial-bounds}
Assume \eqref{eq:kernel-input} and the fixed parameters of
\cref{def:kernel-parameters}. Conditional on every history at its start,
the trial returns with probability greater than $63/64>0.98$.
Every run, whether accepted or rejected, uses at
most $6Ns^{24}$ queries.
\end{lemma}
\begin{proof}
For the success probability, every check passes outside the failure
events bounded in \cref{lem:app-kernel-closure}.
The five contributions---reserve mass, reserve cell
count, selected parents, assignment separation and payment errors---sum
to less than $1/64$. Each bound accounts for the random choices in the
relevant steps, either directly or by fixing earlier choices and then
averaging over them. Since the trial uses fresh randomness, the same
guarantee holds whatever happened before it.

For the query bound, only new endpoints require queries. The grid, the
samples and the corrections create at most $k/\eta$, $m$ and $2k$ new
endpoints respectively. Since $k/\eta\le m$ and $2k\le m$, their total
is at most $3m$. Each costs at most one $\Cut$ and $N$ $\Eval$ queries,
hence at most $6Nm=6Ns^{24}$ queries altogether.
\end{proof}

Together with \cref{prop:kernel-output} and the deterministic empty-buffer
case of \cref{sec:overview}, this proves \cref{thm:kernel}.

\subsection{Partial allocation from nearly equal pieces}\label{sec:kernel-consequences}

We now use Repair to allocate cake from scratch. We need pieces whose
values are close to prescribed fractions for every agent at once.
The following elementary construction
uses the same fractional-plan simplification as the reserve payments.
Whole cells can be assigned directly; discarding the few split cells
makes each target an upper bound and limits the shortfall. This is the
one-sided simultaneous carving used in shrinking; compare \cite[Lemma~4.6]{YeBai2026}.

\begin{lemma}[Simultaneous carving]\label{lem:completion-carving}
For a finite union $U$ with recorded endpoint values, $k$ active agents,
$M\ge1$ target fractions $f_1,\ldots,f_M\ge0$ with $\sum_jf_j\le1$,
and accuracy $0<\alpha<1$, one can deterministically construct disjoint
pieces $A_j\subseteq U$ satisfying
\[
 (f_j-\alpha)v_i(U)\le v_i(A_j)\le f_jv_i(U)
 \quad\text{for every active }i\text{ and every }j.
\]
The cost is at most $(N+1)(k^2M/\alpha+k)$ queries; all unused cake stays
in the residue.
\end{lemma}
\begin{proof}
If $U$ contains no interval of positive length, all target values vanish
and empty pieces suffice.
Otherwise hungriness gives $v_i(U)>0$ for every agent.

\emph{Small cells.} We cut the ordered union $U$ at the multiples of
$\alpha v_i(U)/(kM)$ in each agent's value,
and record endpoint values for all $N$ original agents. Every cell is
then worth at most that amount to every agent. Each of the $k$ agents
contributes at most $kM/\alpha+1$ points, so at most $k^2M/\alpha+k$ new
points are needed, each costing at most one cut and $N$ evaluation
queries (\cref{lem:union-cut}).

\emph{Rounding the plan.} The numerical plan that gives label $j$ a
$f_j$ share of each cell, leaving the rest unallocated, meets all $kM$
value targets. By \cref{lem:sparse-fractions}, we may change it,
preserving those targets, so that at most $kM$ cells are split, that
is, have more than one positive fraction (unallocated cake counting as
a label). A cell with a single positive fraction has fraction one, and
we give it whole to its label or leave it unallocated; the split cells
stay in the residue. The residual identity \eqref{eq:rounding-residual}
says that each assigned piece gains nothing and loses at most
$kM\cdot\alpha v_i(U)/(kM)=\alpha v_i(U)$. Both bounds use only
recorded values.
\end{proof}

\subsection{Repeated shrinking}\label{sec:kernel-shrink}

We can set aside almost all the cake in nearly equal pieces and keep
only a small buffer for repair. Once those pieces have been allocated,
the cake left over is already small. Finer carving costs more queries,
but still fits the allowance for a repair trial. We use a reduction
factor $s^8$ per successful step to prove \cref{lem:shrink}.

\begin{proof}[Proof of \cref{lem:shrink}]
\emph{One attempt.} Let $U$ be the cake remaining at its start.
If $U$ contains no interval of positive length, there is nothing left
of value to allocate; with one participant, give her all of $U$.
Otherwise every participant values $U$ positively. Apply
\cref{lem:completion-carving} with $k$ equal target fractions and accuracy
\[
 f=\frac{1-1/(2s^8)}{k},\qquad \alpha=\frac{\tau}{4s^8}.
\]
Write $W_j$ for the resulting endowments and
$B=U\setminus\bigcup_jW_j$ for the repair buffer. The carving bounds give
\[
\begin{gathered}
 (f-\alpha)v_i(U)\le v_i(W_j)\le f v_i(U),\\[2pt]
 \frac{v_i(U)}{2s^8}\le v_i(B)
 \le\left(\frac1{2s^8}+\frac{k\tau}{4s^8}\right)v_i(U)
 \le\frac{v_i(U)}{s^8}.
\end{gathered}
\]
The last inequality uses $k\tau\le2$, immediate from
$\tau=1/(4s^{13})$ and $k\le s$. The endowment values differ by at most
$\alpha v_i(U)\le(\tau/2)v_i(B)$, so the repair condition
\eqref{eq:kernel-input} holds, and the buffer has positive value.
Run one fresh repair trial. A rejection ends this attempt without
allocating cake. On success all endowments remain allocated whole, so
the new residue lies in $B$ and is worth at most $v_i(U)/s^8$ to every
participant. Check this bound and envy-freeness from the recorded values
before accepting the step.

\emph{Repeated steps.} Let $G\ge1$ be the requested total reduction.
For $G=1$, return the empty addition. Otherwise find the least integer
$r$ with $(s^8)^r\ge G$ by repeated multiplication. Retry on the current
residue until one attempt succeeds, add its allocation to $Y$, and
continue until $r$ steps have succeeded. Stop earlier if no cake remains.
A rejected attempt starts afresh from the same input; its recorded
endpoints do not become boundaries in the next trial's sampling grid.

Each successful step reduces every remaining value by a factor at least
$s^8$. After $r$ steps it is therefore at most $s^{-8r}\le1/G$ times
its initial value. Adding disjoint envy-free allocations adds nonnegative
pairwise advantages and cannot reduce earlier margins. Before returning,
check disjointness, envy-freeness and the final residue bound exactly.

\emph{Queries and attempts.} The finer carving costs at most
\[
 (N+1)\left(\frac{4k^3s^8}{\tau}+k\right)
 =(N+1)(16k^3s^{21}+k)=O(Ns^{24}),
\]
including endpoint values for all $N$ original agents. Thus it fits the
same allowance as repair. Conditional on any history at the start of an
attempt, carving supplies valid inputs and fresh repair succeeds with
probability greater than $63/64$. By \cref{lem:retry}, each successful
step needs at most $64/63$ attempts in conditional expectation. There
are at most $r=O(1+\log G/\log s)$ steps. Summing gives the claimed
attempt and query bounds. Each attempt includes both carving and repair,
whether accepted or rejected.
\end{proof}
\Needspace{10\baselineskip}
\section{Advantages relative to the remaining cake}\label{sec:snapshot}

Repeated partial allocation shrinks the residue but need not exhaust it.
To let an agent stop, her advantage over those who continue must cover
her value of the remaining cake. The margins of \cref{sec:kernel} may be
tiny compared with the residue, because they are proportional to observed
disagreement. Here we obtain an advantage that is an inverse-polynomial
fraction of its owner's residue value; polynomial shrinking can then turn
it into domination.

With large observed disagreement, two small pieces supply the preference
that repair will preserve. With small
positive disagreement, we make the residue small to match: first allocate
most of the cake, keeping only a repair buffer, so that even a small margin
covers a fixed fraction of the residue. With no observed disagreement, we
try an equal division.

Our goal is \cref{thm:snapshot}: an envy-free allocation with an
advantage at least $\theta=\tau/(768m)$ times its owner's value of the
residue.

\subsection{Observed disagreement}\label{sec:snapshot-order}

The input is a finite union $U$ and $k$ agents. If $U$ contains no
interval of positive length, give it to one agent and return: everyone
values this cake at zero. Otherwise hungriness gives $v_j(U)>0$ for
every participant. With at most three agents, use
\cref{lem:base-cases}.

Henceforth $k\ge4$. Index the participants by $[k]$ and divide each
agent's values by her value of $U$, so that $v_j(U)=1$.
This changes neither envy-freeness nor an advantage measured relative
to the remaining cake. Fix the least-indexed agent $c$ as the reference
agent before drawing the value table. In the small-disagreement
construction she supplies the cut values, so we call her the \emph{cutter}.

Each attempt constructs one set of starting pieces and repairs them if needed.
No cake is assigned unless the whole attempt succeeds.
Both cases with positive disagreement use the same repair procedure.

\Needspace{8\baselineskip}
\smallskip\noindent\textbf{One snapshot attempt.}
Use the fixed mesh $\eta_E=s^{-23}$ and repair tolerance
$\tau=1/(4s^{13})$ of \cref{def:snapshot-parameters}.
\begin{enumerate}[label=(\roman*)]
\item \emph{Value table.} Form the coarse grid with mesh $\eta_E$.
  Draw $m=s^{24}$ independent sample points from
  $\mu=k^{-1}\sum_{j=1}^k v_j|_U$, refine the grid at them and record
  all new endpoint values for the $N$ original agents. No reserve labels
  are drawn at this stage. The coarse cell count satisfies
  $M_0\le2k/\eta_E\le m$, so there are at most $M_0+m\le2m$
  refined cells, each worth at most $\eta_E$ to every participant.
\item \emph{Observed disagreement.} Compare every agent with the fixed
  cutter:
  \begin{equation}\label{eq:snapshot-observed}
   D_j=\sum_{E'\text{ refined cell}}|v_j(E')-v_c(E')|,
   \qquad D=D_{i^*}=\max_{j\ne c}D_j.
  \end{equation}
  The absolute values prevent opposite preferences from cancelling.
  Choose a maximising agent $i^*\ne c$ by a fixed rule, also when $D=0$.
  Thus $D$ bounds each agent's disagreement with the cutter on this
  table. It need not bound disagreement on arbitrary subsets inside a
  cell; the sampling estimate below controls the pieces we construct.
\item \emph{Starting pieces.} If $0<D<\tau/16$, construct $k$ pieces
  equal to the cutter and a separate buffer, as in
  \cref{sec:snapshot-near}. Reject if their actual values fail the
  checks there. If $D=0$, use one circular start to partition $U$
  equally for the cutter; return if the partition is exactly envy-free,
  and reject otherwise.
  If $D\ge\tau/16$, use the two cells with opposite value differences
  and the empty endowments of \cref{sec:snapshot-far}; their repair input
  condition holds deterministically.
\item \emph{Fresh repair.} For either nonzero branch, run one repair
  trial on these endowments and buffer. It uses its own sample
  points, reserve labels, random thresholds and permutation for the seed
  pieces, and correction starts. Return only if the
  resulting allocation is envy-free and is either complete or satisfies
  \eqref{eq:snapshot-certificate}, checked in actual values; otherwise
  reject the snapshot attempt.
\end{enumerate}
A rejected attempt assigns no cake and discards all its temporary data.
Repeat all four steps on the same input $U$ with fresh randomness.
Each attempt uses at most one repair trial; failure requires a fresh
value table. The small positive-disagreement construction uses two fresh
circular starts after that table is fixed.

The threshold $\tau/16$ comes from the repair input condition
\eqref{eq:kernel-input}. We ensure that every agent's values of the
starting pieces differ by at most $\tau$ times her buffer value, so any
assignment meets that condition. For small disagreement the buffer has
cutter-value $8D/\tau<1/2$: it is large enough to cover the differences
but small enough to leave a useful initial allocation. For larger
disagreement, two recorded cells already provide a suitable input.

\subsection{Preference differences under repair}

Repair can reassign the starting pieces and add compensation, so a
preference for one starting piece need not become an advantage in the
final allocation. Instead, we compare how strongly two agents prefer the
same pair of pieces. We call this difference a \emph{contrast}. For
example, in reserve units, if one agent prefers the first piece by five
units and another prefers it by two, their contrast is three.
For agents $a,b$ and pieces $E,F$, the contrast in reserve-normalised
values $w$ is
\[
 \bigl(w_a(E)-w_a(F)\bigr)-\bigl(w_b(E)-w_b(F)\bigr)
 =(w_a-w_b)(E)-(w_a-w_b)(F).
\]

There are two reasons this helps. First, swapping the random additions
to the two pieces reverses their contribution to the contrast. The two
orders are equally likely, so the additions preserve the original
contrast with probability at least one half. Second, after each agent's
values are divided by her own reserve value, a piece's ideal numerical
credit is the same in every row. These ideal credits cancel when we
subtract the two agents' preferences. If the contrast remains large,
some final preference must therefore remain large too. The payment
bounds make this last statement quantitative.

The following lemma converts a contrast into an advantage over the
residue. Repair's \emph{preparation} fixes the reserve values, so the
selected agents and pieces may depend on them. Selection precedes the
random seed thresholds and permutation, preserving the symmetry used
below. The lower bound $\Lambda/q$ must hold after every successful
preparation.

\begin{lemma}[A preference difference relative to the residue]\label{lem:snapshot-common-repair}
Fix a valid repair input and its preceding history, with at least two
agents and a buffer $B$ that every agent values positively. Suppose that
after each successful preparation, two agents and two endowments can be
chosen whose contrast, measured in the reserve-normalised values $w_i$,
is at least $\Lambda/q$, for a fixed $\Lambda>0$. Make this choice before
the random thresholds and permutation. With probability at least
$31/64$, repair returns an envy-free allocation $X$ with residue
$T$ and agents $a,b$ such that
\[
 v_a(X_a)-v_a(X_b)\ge \frac{\Lambda}{8}v_a(T).
\]
\end{lemma}
\begin{proof}
\emph{The random additions.}
Fix a successful preparation and order the selected agents $i,r$ and
endowments $W_u,W_v$ so that
\[
 (w_i-w_r)(W_u)-(w_i-w_r)(W_v)\ge\Lambda/q.
\]
If $Z_u,Z_v$ are the seed pieces assigned to these columns, their
contribution is
$(w_i-w_r)(Z_u)-(w_i-w_r)(Z_v)$.
Swapping the columns preserves the conditional distribution of the seed
pieces (\cref{lem:kernel-symmetry}) and changes the sign of this
contribution. It is therefore nonnegative with probability at least
$1/2$, including ties.

Let $P$ be successful preparation, $S$ preservation of the selected
contrast, and $G$ a return of the repair trial. Every return passed
preparation, so $G\subseteq P$. The repair failure bound
$\Pr(G^{\rm c})\le1/64$ in \cref{lem:kernel-trial-bounds} gives
\[
 \Pr(S\cap G)\ge\tfrac12\Pr(P)-\Pr(P\cap G^{\rm c})
 \ge\tfrac12-\Pr(G^{\rm c})\ge\tfrac{31}{64}.
\]
This is a joint bound over the whole trial; it does not assume that repair
is likely to return after every selected reserve or table.

\emph{The final margin.}
On $S\cap G$, let $d_{ij}=w_i(X_i)-w_i(X_j)\ge0$ and
$\Delta_{\max}=\max_{i,j}d_{ij}$.
Payment accuracy and the final envy margin
(\cref{eq:kernel-representation,prop:kernel-output}) give
\[
 |e_{ij}|\le\lambda A_{ij}\le d_{ij}.
\]
Thus the credited values in row $i$ satisfy
\[
 V_{i,\pi(j)}+t_j=w_i(X_i)-(d_{ij}+e_{ij}),
 \qquad 0\le d_{ij}+e_{ij}\le2\Delta_{\max}.
\]
Each row therefore has range at most $2\Delta_{\max}$. Ideal column
credits cancel between two rows' preferences, so any surviving contrast
is at most $4\Delta_{\max}$. Hence $\Delta_{\max}\ge\Lambda/(4q)$.

Choose $a,b$ attaining $\Delta_{\max}$. The reserve mass check
\eqref{eq:kernel-reserve-mass} gives
$v_a(R)\ge(q/2)v_a(B)$. Every endowment is assigned whole, so all
remaining cake lies in $B$. Returning to actual values therefore gives
\[
 v_a(X_a)-v_a(X_b)=v_a(R)\Delta_{\max}
 \ge\frac{\Lambda}{8}v_a(B)
 \ge\frac{\Lambda}{8}v_a(T).
\]
In particular, the argument needs no comparison between different
agents' reserve values.
\end{proof}

\subsection{Small disagreement and the repair buffer}\label{sec:snapshot-near}

When $D=0$, the agents agree on every recorded cell. They may still
disagree inside those cells, so an equal division for the cutter needs
checking. One fresh circular start partitions $U$ into $k$ pieces of
cutter-value $1/k$ (\cref{lem:sampling-crop}). Record the new endpoints
for all $N$ original agents and return only if this complete allocation
is envy-free. The sampling argument below shows that these checks pass
outside its stated failure event.

Now assume $0<D<\tau/16$. We seek pieces $W_1,\ldots,W_k$ that the
cutter values equally, together with a buffer $B$ and a preference
between two pieces for agent $i^*$.
Pieces nearly equal for every agent have classical precedents in
Robertson--Webb's near-exact division~\cite[Theorem~3.2]{RobertsonWebb1997}.
Robertson and Webb's proof of their Theorem~4.1 also gives pieces that one designated cutter
values exactly equally; see also Pikhurko's partition
lemma~\cite[Lemma~1]{Pikhurko2000}.
The construction here also preserves a measured preference and controls
the buffer size and the distribution of new endpoints.
A repair trial will assign these pieces whole and add buffer cake to
remove envy, so that only buffer cake can remain unallocated.

Write $K=8/\tau$ for the buffer multiplier. The buffer has cutter-value
$KD<1/2$, and each starting piece has cutter-value $t=(1-KD)/k$. We cut the
pieces at these final sizes, so none is later trimmed to make room for
the buffer.

\smallskip\noindent\textbf{Values of constructed pieces.}
The pieces built in the zero and small-disagreement cases may have
boundaries inside refined cells, where the table does not determine how much two
agents' values differ. We therefore compare agents through the cell
estimate $\widehat\sigma_{jc}$ of $v_j-v_c$. Within each cell it joins the
recorded cumulative values by a straight line: a part occupying one tenth
of the cell's length receives one tenth of its recorded difference
(\cref{def:sampling-model}). This estimates values between recorded
boundaries; it does not assume that actual valuations are uniform. Its
positive and negative masses balance, because both agents value $U$ at
one, and their total absolute mass, or total variation, is $D_j$;
consequently $|\widehat\sigma_{jc}(E)|\le D_j/2$ for every set $E$.
Set $\varepsilon=s^{-2}$. The sampling estimate gives, simultaneously
on every piece of the zero-case partition or small-disagreement
construction,
\[
 |(v_j-v_c)(E)-\widehat\sigma_{jc}(E)|\le \varepsilon D_j,
\]
except on an event $\mathcal B_{\rm geo}$ of probability at most
$\varepsilon$ (\eqref{eq:snapshot-geo-failure} in
\cref{sec:app-snapshot-menu}). The bound averages over both the sample
points and the later cuts; it is not a guarantee for every observed table.
Thus $\varepsilon$ serves as both the
relative sampling accuracy and the failure allowance
(\cref{def:snapshot-parameters}).

When $D>0$, the first piece $W_1$ favours cells that $i^*$ values more than the cutter,
giving a positive expected difference. The remaining cake has the
opposite difference, and on average a random partition passes a
proportional share of it to $W_2$.
The preference is of order $D/k$, while the buffer has value of order
$D/\tau$. Their ratio is an inverse-polynomial fraction, independent of
how small $D$ is. Random starts also control the new
endpoints, allowing the sampling estimate to transfer the comparison to
actual values.

\smallskip\noindent\textbf{The starting pieces and buffer.}
Write $\widehat\sigma=\widehat\sigma_{i^*c}$, let $S_+$ be the union of
refined cells with positive $\widehat\sigma$-mass, and put $u=v_c(S_+)$.
\begin{enumerate}[label=(\arabic*)]
\item \emph{The first piece.} If $u\ge t$, take $W_1$ to be a circular
  piece of cutter-value $t$ from $S_+$. If $u<t$, retain all of $S_+$ and
  fill the shortfall with a circular piece of cutter-value $t-u$ from its
  complement. The piece uses a fresh uniform starting value.
\item \emph{The remaining pieces and buffer.} With one further fresh
  circular start, partition $U\setminus W_1$ into $W_2,\ldots,W_k$, each
  of cutter-value $t$, and $B$ of cutter-value $KD$; the required values
  sum to $1-t$.
\item \emph{Recorded values.} Record every new endpoint's value for
  all $N$ original agents. These values determine whether the partition
  passes \eqref{eq:snapshot-menu-checks}. Keep the original cell estimate
  unchanged when recording these endpoints.
\end{enumerate}
All these cuts follow the ordered components of the cake, skipping the
gaps between them, as in \cref{lem:sampling-crop}.

The next lemma gives a modest chance of a useful preference and bounds
the new endpoints so that the sampling estimate applies.

\Needspace{11\baselineskip}
\begin{lemma}[Equal pieces with a repair buffer]\label{lem:snapshot-menu}
Conditional on every table with $0<D<\tau/16$ and the preceding history,
the construction has probability at least $1/(8k)$ of producing an estimated preference
$\widehat\sigma_{i^*c}(W_1)-\widehat\sigma_{i^*c}(W_2)\ge D/(8k)$.
On this event and off $\mathcal B_{\rm geo}$, the partition
$U=B\,\dot\cup\,W_1\,\dot\cup\cdots\dot\cup W_k$ satisfies
\begin{equation}\label{eq:snapshot-menu-checks}
\begin{gathered}
 v_c(B)=KD,\qquad v_c(W_b)=(1-KD)/k\quad(b\in[k]),\\
 v_{i^*}(W_1)-v_{i^*}(W_2)\ge D/(32k),\qquad
 (K-2)D\le v_j(B)\le(K+2)D,\\
 \osc_b v_j(W_b)\le4D\le\tau v_j(B)\quad(j\in[k]).
\end{gathered}
\end{equation}
The buffer bounds hold for every $j\in[k]$; $\osc_b$ is the largest value
minus the smallest over all starting pieces.
The construction exposes at most $k+2$ new endpoints, each with conditional distribution at most
$2k^2\mu$ before its group's shared randomness is drawn.
\end{lemma}

The endpoints of one circular piece or partition share its random start.
The endpoint bound says that,
with earlier choices fixed and before that start is drawn, each new
endpoint lands in a region $E$ with probability at most $2k^2\mu(E)$;
endpoints from the same start need not be independent. Together with the endpoint count, these
conditional bounds give the grouped condition in
\cref{def:sampling-slots}.

\begin{proof}
Fix the snapshot table and preceding history. We first obtain an
estimated preference from the opposite expected differences on $W_1$
and $W_2$, then use the sampling estimate to control actual values.

\emph{A preference in the cell estimate.}
The two agents value $U$ equally, so the positive and negative masses of
$\widehat\sigma$ balance:
\[
 \widehat\sigma(S_+)=D/2=-\widehat\sigma(U\setminus S_+).
\]
The buffer has cutter-value less than one half, so $1/(2k)<t\le1/k$.
The point-inclusion identity of \cref{lem:sampling-crop} gives
\[
 \E\widehat\sigma(W_1)=
 \begin{cases}
  tD/(2u),&u\ge t,\\
  D(1-t)/(2(1-u)),&u<t.
 \end{cases}
\]
In either case the expectation is at least $tD/2$: the first piece favours
the positive cells. The remainder has signed value
$-\widehat\sigma(W_1)$. Its fresh random partition therefore gives
\[
 \E[\widehat\sigma(W_2)\mid W_1]
 =-\frac{t}{1-t}\widehat\sigma(W_1).
\]
Here conditioning on $W_1$ means fixing the entire first-step history.
For $\chi=\widehat\sigma(W_1)-\widehat\sigma(W_2)$, it follows that
\[
 \E\chi=\frac{\E\widehat\sigma(W_1)}{1-t}\ge\frac D{4k},
\]
using $t>1/(2k)$. On the other hand $\chi\le D$, because the total
absolute mass of $\widehat\sigma$ is $D$. Splitting the expectation
according to whether $\chi\ge D/(8k)$, and bounding $\chi$ by $D$ on that
event and by $D/(8k)$ off it (where $\chi$ may be negative), gives
\[
 \E\chi\le \frac D{8k}
       +D\Pr\!\left[\chi\ge\frac D{8k}\right].
\]
Hence the estimated preference occurs with probability at least $1/(8k)$.

\emph{Actual values.}
The cutter-values in the first line of \eqref{eq:snapshot-menu-checks}
hold by construction. Since the cutter values $W_1$ and $W_2$ equally,
$v_{i^*}(W_1)-v_{i^*}(W_2)$ is the difference of the values of $v_{i^*}-v_c$ on the
two pieces, and outside $\mathcal B_{\rm geo}$ each of these values is
within $\varepsilon D$ of the cell estimate. The preference therefore loses at
most $2\varepsilon D$ and keeps at least $D/(16k)$, since
$2\varepsilon\le1/(16k)$ (\cref{sec:app-snapshot-menu}); this is more
than the check requires. For the remaining checks we compare every agent
with the cutter. The observed disagreement of $j$ and $c$ is at most $D$,
and their estimated difference has total mass zero, so it is at most half
its total variation on any set. Together with the sampling estimate this
gives, for every agent $j$ and every constructed piece $E$,
\[
 |v_j(E)-v_c(E)|\le(1/2+\varepsilon)D.
\]
For the buffer, $v_j(B)$ is therefore within $(1/2+\varepsilon)D\le2D$
of $v_c(B)=KD$. Two starting pieces have equal cutter-values, so their values
to $j$ differ by at most twice that bound, hence by at most $4D$.
Finally, $\tau v_j(B)\ge\tau(K-2)D\ge4D$ because $K-2\ge4/\tau$.

\emph{Endpoint distributions.}
By \cref{lem:sampling-crop}(c), an endpoint's distribution is at most the owner
measure divided by a lower bound on the cutter-value of the domain being
cut, so we bound these domains from below. The circular piece in step (1) adds at
most two endpoints in a domain of cutter-value at least $t>1/(2k)$ when
$u\ge t$, or at least $1-t\ge3/4$ when $u<t$; the partition in step (2)
adds at most $k$ endpoints in a domain of cutter-value $1-t$, and
boundaries already used in $W_1$ are counted only once. Since
$v_c\le k\mu$, every new endpoint has conditional distribution at most $2k^2\mu$
before its group's shared start.
\end{proof}

For small disagreement, the construction makes both the initial preference and
the buffer value proportional to $D$. Their ratio supplies the fixed
contrast bound required by \cref{lem:snapshot-common-repair}.

\begin{lemma}[Small-disagreement repair]\label{lem:snapshot-near}
For $0<D<\tau/16$, given starting pieces passing
\eqref{eq:snapshot-menu-checks}, one
fresh repair returns, with probability at least $31/64$, an envy-free
allocation with residue $T\subseteq B$ and some margin
\[
 v_a(X_a)-v_a(X_b)
 \ge \frac{v_a(T)}{384k(K+2)}\ge\theta v_a(T).
\]
\end{lemma}
\begin{proof}
The checks make every buffer value positive and imply the repair
input condition. Fix any successful preparation, and write
$r_{i^*}=v_{i^*}(R)$ for agent $i^*$'s reserve value. The cutter values $W_1,W_2$
equally, so the contrast between rows $i^*,c$ is
\[
 (w_{i^*}-w_c)(W_1)-(w_{i^*}-w_c)(W_2)
 =\frac{v_{i^*}(W_1)-v_{i^*}(W_2)}{r_{i^*}}
 \ge\frac{D}{32kr_{i^*}}.
\]
The reserve and starting-piece checks give
$r_{i^*}\le(3q/2)(K+2)D$. Thus the contrast is at least
$1/[48kq(K+2)]$ after every successful preparation. Apply
\cref{lem:snapshot-common-repair} with
$\Lambda=1/[48k(K+2)]$. The resulting margin has the stated size, and
$1/[384k(K+2)]\ge\theta$ by \cref{lem:snapshot-scales}.
\end{proof}

\subsection{Large disagreement and two cells}\label{sec:snapshot-far}

When $D\ge\tau/16$, two tiny recorded cells suffice as endowments;
all other endowments are empty. The remaining cake is a large repair
buffer. The two agents both value the whole input at one, so their
positive cell differences add to $D/2$, and their negative differences
add to $-D/2$. Among the $M\le2m$ refined cells, choose $F$ with the
largest difference $v_{i^*}(F)-v_c(F)$ and $G$ with the smallest, breaking
ties by a fixed rule. Then
\[
 v_{i^*}(F)-v_c(F)\ge D/(2M),\qquad
 v_c(G)-v_{i^*}(G)\ge D/(2M).
\]
Thus $F$ and $G$ are distinct: $i^*$ values $F$ more than the cutter
does, whereas the cutter values $G$ more than $i^*$ does. Set
\[
 W_1=F,\qquad W_2=G,\qquad W_3=\cdots=W_k=\emptyset,
 \qquad B=U\setminus(F\cup G).
\]
Every agent gives $B$ value at least $1-2\eta_E\ge1/2$ and has endowment
range at most $\eta_E\le\tau/2\le\tau v_j(B)$. Thus the repair input
condition holds deterministically. These pieces use recorded boundaries
and require no further cuts.

Repair divides each agent's values by her own reserve value. Whichever
reserve value is smaller, one of the two cells retains a signed
difference of the required size. Comparing that cell with an empty
endowment, whose difference is zero, gives the contrast needed below.

\begin{lemma}[Large-disagreement repair]\label{lem:snapshot-far}
For the two-cell input above, a fresh repair returns, with probability at
least $31/64$, an envy-free allocation with residue $T\subseteq B$
and some margin
\[
 v_a(X_a)-v_a(X_b)\ge\frac{D}{24M}v_a(T)\ge\theta v_a(T).
\]
\end{lemma}
\begin{proof}
Fix any successful preparation, let $r_j=v_j(R)$, and put
$\gamma=v_{i^*}/r_{i^*}-v_c/r_c$. If $r_{i^*}\le r_c$, the preference on $F$
survives normalisation:
\[
 \gamma(F)=\frac{v_{i^*}(F)-v_c(F)}{r_{i^*}}
       +v_c(F)\left(\frac1{r_{i^*}}-\frac1{r_c}\right)
 \ge\frac{D}{2Mr_{i^*}}.
\]
If $r_c\le r_{i^*}$, the same calculation on $G$ gives
$-\gamma(G)\ge D/(2Mr_c)$. The reserve check gives
$r_j\le(3q/2)v_j(B)\le3q/2$ in either case. Hence one cell has
absolute difference at least $D/(3qM)$. An empty endowment has difference
zero and exists because $k\ge4$. Select the cell and orient its
comparison with this empty endowment before the random thresholds and
permutation. This gives the contrast required by
\cref{lem:snapshot-common-repair}, with $\Lambda=D/(3M)$.

The resulting relative margin is $\Lambda/8=D/(24M)\ge\theta$ by
\cref{lem:snapshot-scales}.
\end{proof}

\subsection{Success probability and query cost}\label{sec:snapshot-success}

An attempt first needs a complete allocation or valid starting pieces
for repair. We bound that probability before drawing the table. Once
the pieces pass their checks, the fresh repair has a constant chance
of preserving a useful advantage. Its bound holds for every checked
input and the full history so far, so the two bounds can be multiplied.

\begin{proposition}[Snapshot attempts and their cost]\label{prop:snapshot-cost}\label{lem:snapshot-success}
For every fixed hungry valuation profile, conditional on
any preceding history, one snapshot attempt with $k$ input agents
returns a complete allocation or the certificate of \cref{thm:snapshot}
with probability at least $1/(64k)$. Every attempt uses
$O(Ns^{24})$ queries. Repeating attempts until one returns terminates
almost surely and costs $O(Nks^{24})$ queries in expectation.
\end{proposition}
\begin{proof}[Proof of the proposition and \cref{thm:snapshot}]
First suppose $U$ has positive length and $k\ge4$.
Let $A$ be the estimated preference event
for a small positive-disagreement table, and let $A$ occur automatically
at zero or large disagreement. By \cref{lem:snapshot-menu},
$\Pr(A\mid\text{table})\ge1/(8k)$ for every table.

Off the joint sampling exception $\mathcal B_{\rm geo}$, an estimated
preference passes the actual-value checks. At $D=0$, every cell estimate
is zero, so every agent values each piece of the equal partition at
$1/k$ and its exact checks pass. The large-disagreement input is valid
deterministically. The exception has probability less than
$\varepsilon$ before the table is drawn (\cref{sec:app-snapshot-menu});
it is not a bound conditional on an arbitrary table. Subtracting it once
therefore gives
\[
 \Pr(\text{complete return or valid repair input})
 \ge\frac1{8k}-\varepsilon\ge\frac1{16k}.
\]

For every checked nonzero input, fix the entire history before its one
fresh repair. \Cref{lem:snapshot-near,lem:snapshot-far} give conditional
acceptance probability at least $31/64$. A complete zero-case return
needs no repair. Consequently
\[
 \Pr(\text{snapshot attempt returns})
 \ge\frac1{16k}\frac{31}{64}\ge\frac1{64k}.
\]
This multiplication uses a conditional bound for every checked input;
it does not assume independence of input selection and repair outcome.

The table, starting pieces and the one repair together use
$O(Ns^{24})$ queries on every attempt, as detailed in
\cref{sec:app-snapshot-cost}. On rejection, discard the temporary data
and leave the input cake unchanged. The same success bound holds after
any such history, so \cref{lem:retry} gives at most $64k$ attempts in
expectation and almost-sure termination. Every return is checked exactly
for envy-freeness and either completeness or the required advantage.

If $U$ has no positive-length component, the immediate return is complete
and envy-free. With at most three agents, the complete base cases succeed
certainly using $O(N)$ queries. Finally, undoing each agent's normalisation
multiplies both sides of her certificate inequality by the same positive
number. This proves \cref{thm:snapshot} for the original values.
\end{proof}
\section{Query cost and computation}\label{sec:accounting}\label{sec:app-parameters}

We bound the cost of each attempt and the number of attempts needed.
Successful attempts add lasting records; rejected attempts discard their
temporary data. Bounding successful attempts therefore controls storage,
while conditional success probabilities control the expected total cost.

Every attempt costs $O(Ns^{24})$ queries, including a failed one
(\cref{lem:kernel-trial-bounds,prop:snapshot-cost,lem:shrink}). The
sampled tables have $O(s^{24})$ new endpoints, and recording each costs
at most one cut and $N$ evaluations. A snapshot uses at most two such
tables and $O(k)$ further endpoints. The carving
inside a shrinking attempt adds $O(k^3s^{21})$ endpoints, which is
$O(s^{24})$ since $k\le s$.

A repair succeeds with conditional probability greater than $1/2$,
and a snapshot attempt with probability at least $1/(64k)$. Including
retries, obtaining a repair therefore costs $O(Ns^{24})$ expected queries,
and obtaining a snapshot costs $O(Nks^{24})$.
Each successful shrinking attempt reduces every participant's residue
value by a factor of at most $s^{-8}$, so $\operatorname{Shrink}(G)$ costs
$O(Ns^{24}(1+\log G/\log s))$ expected queries. A failed inner repair ends
the entire enclosing attempt; all these bounds include its preparation.

The common shrinking factor satisfies $2/\theta=6144s^{37}\le s^{40}$,
so five successful steps suffice. Each separation factor satisfies
$2k^2/\tau=8k^2s^{13}\le s^{16}$, so two suffice.
Together with \cref{prop:completion-stage}, this gives the table below.
One successful shrinking attempt means one $s^8$-factor step, not an
entire call to $\operatorname{Shrink}(G)$. The last column remains valid
after any preceding history.
\begin{center}
\begin{tabular}{@{}lcc@{}}
\toprule
Task at active size $k$ & Successful attempts & Expected tries per success \\
\midrule
Snapshots & $k^2$ & at most $64k$ \\
Common shrink & at most $5$ & at most $2$ \\
Separation shrinks & fewer than $2k^4$ & at most $2$ \\
Final repair & $1$ & at most $2$ \\
\bottomrule
\end{tabular}
\end{center}

\Needspace{10\baselineskip}
\begin{lemma}[Successful work and remaining attempts]\label{lem:total-trials}
For a sufficiently large fixed integer $C$, let
\[
 B_N=C N^5.
\]
Completion has at most $B_N$ successful attempts on every run.
Initially and at every reached boundary between attempts, its expected
number of further attempts, conditional on the full history, is at
most $B_N$. This integer can be computed using fixed integer arithmetic.
\end{lemma}
\begin{proof}
A run has at most one stage of each active size $k\ge4$, because each
completed stage removes at least one agent. Separation shrinking uses
$O(k^4)$ successful steps at size $k$ and dominates the other
successful counts in the table. Summing over the at most $N$ stage sizes
gives $O(N^5)$ successful attempts. Early exits only reduce this
bound, and the final case of at most three agents is deterministic.

Now fix the full history at any reached attempt boundary. For the
current request, the expected number of further tries is still at most
$64k$ for a snapshot and two for a shrinking step or repair
(\cref{lem:retry}). Earlier failures do not alter these bounds.
A partly completed shrinking call has no more successful steps left
than its original allowance. The current and future requests thus use
at most the counts in the table. Their inputs may depend on earlier
outcomes; apply the waiting bound at each request's starting history
and sum, without requiring independence. Snapshots contribute
$\sum_{k=4}^N O(k^3)=O(N^4)$ expected attempts; separation shrinking
contributes $O(N^5)$ and dominates the other terms. Thus one
absolute choice of $C$ gives both bounds. Computing $B_N$ uses only
fixed integer arithmetic.
\end{proof}

\Needspace{7\baselineskip}
\begin{proposition}[Expected query bound]\label{prop:continuous-queries}
With the numerical choices in
\cref{def:kernel-parameters,def:snapshot-parameters}, completion uses
$O(N^{30})$ expected queries. This bound holds for every hungry
valuation profile and after every history at which the call's input
conditions hold. The expectation is over the call's fresh random choices.
\end{proposition}
\begin{proof}
Since $s=\Theta(N)$, every attempt costs $O(N^{25})$ queries.
By \cref{lem:total-trials}, the expected number of attempts is
$O(N^5)$. Their product is $O(N^{30})$. The initial recordings and final small-agent
case cost $O(N)$ queries. All remaining decisions use recorded values
and introduce no queries, so they do not change the bound.
\end{proof}

For computation, a rejected attempt restores the working state at its
start. Only the at most $B_N$ successful attempts add lasting records.
The current attempt adds at most $O(s^{24})$ temporary endpoints, so
the working data have polynomial size on every run. The algorithms in
\cref{app:computation} then give polynomial work per attempt, including
restoration, and hence polynomial expected total work.

The remaining-attempt bound also controls long runs. By Markov's
inequality, after any reached attempt boundary the chance of needing
more than another $2B_N$ attempts is at most $1/2$. Each further block
therefore halves the probability of still running, without discarding
earlier successes (\cref{lem:continuation-cap}).

\Needspace{6\baselineskip}
\begin{proposition}[Bounded work with reported failure]\label{prop:continuous-capped}
There is a protocol that, for every hungry valuation profile of $N$
agents and every $0<\delta<1$, uses
$O(N^{30}(1+\log(1/\delta)))$ queries and
$N^{O(1)}(1+\log(1/\delta))$ operations on every run.
It returns a complete, exactly envy-free allocation with probability
at least $1-\delta$ and otherwise reports failure. Every returned
allocation is correct. Operations are counted in
\cref{def:computation-model}.
\end{proposition}
\begin{proof}[Proof of \cref{thm:main,prop:main-computation,prop:continuous-capped}]
After recording all agents' total values, the uncapped procedure runs
\cref{thm:completion} on $[0,1]$. Its local procedures terminate almost
surely by \cref{thm:kernel,thm:snapshot,lem:shrink,lem:retry}.
Completion makes finitely many such calls on decreasing agent sets and
checks final coverage and envy inequalities exactly.

The query bound follows from \cref{prop:continuous-queries}, and
\cref{prop:polynomial-work} proves the expected operation bound.
For the confidence version, find the least positive integer $r$ with
$2^{-r}\le\delta$ by repeated halving. Allow at most $2rB_N$
attempts in the same execution, reporting failure only if another is
needed. Allow all deterministic processing and final checks after the
last permitted attempt. \Cref{lem:continuation-cap} proves the success
guarantee; multiplying the per-attempt query and operation bounds by
$2rB_N$ proves the cost caps. No incorrect allocation is returned.

Taking $N=n$ gives the stated bounds in
\cref{thm:main,prop:main-computation}.
\end{proof}

\Needspace{20\baselineskip}
\subsection{Nearly complete partial allocations}

If a small residue is acceptable, shrinking alone suffices: it gives
every agent almost a proportional share without the completion stages.
The same confidence argument supplies a fixed query budget.

\begin{maintheorem}[Nearly complete partial envy-freeness]\label{thm:partial}
There is a randomised Robertson--Webb protocol for every hungry
valuation profile of $n\ge1$ agents in \cref{sec:model}. Given
$\delta\in(0,1)$, it uses at most
\[
 O\!\left(n^{25}(1+\log(1/\delta))\right)
\]
queries on every run, with a bound independent of the valuations.
It returns an allocation with probability
at least $1-\delta$; otherwise it reports failure. Every returned
allocation is exactly envy-free, has finite-union shares and leaves a
residue $T$ satisfying
\[
 v_i(T)\le n^{-2}v_i([0,1]),\qquad
 v_i(X_i)\ge\frac{1-n^{-2}}{n}v_i([0,1])
 \quad(i\in[n]).
\]
For one agent, the whole cake is allocated deterministically.
\end{maintheorem}
\begin{proof}
For $n\ge2$, record the total values and run $\operatorname{Shrink}(n^2)$
with $N=k=n$. By \cref{lem:shrink}, the residue has the stated value.
Envy-freeness gives
$n v_i(X_i)\ge\sum_jv_i(X_j)=v_i([0,1])-v_i(T)$, proving the share.

Since $s^8\ge n^2$, one successful shrinking step suffices.
By the proof of \cref{lem:shrink}, after every reached attempt boundary
it needs at most $64/63$ further attempts in conditional expectation.
The uncapped expected query cost is therefore $O(n^{25})$.
By \cref{lem:continuation-cap}, allowing $O(1+\log(1/\delta))$ attempts
gives failure probability at most $\delta$, with final deterministic
checks after the last permitted attempt. Each attempt costs $O(n^{25})$
queries, proving the cap, including the initial recordings. All returned allocations are checked
exactly. The one-agent case is immediate.
\end{proof}
\section{Further questions}\label{sec:conclusion}

Two questions remain open:
can a deterministic protocol achieve a polynomial query bound for every
instance, and can fair coins replace the uniform-real draws with an
expected number of bits bounded by a polynomial in the number of agents
alone?
Even a polynomial random-bit bound would leave a separate question about
binary running time: our model allows arbitrary real query answers and
exact real operations at unit cost.
\appendix
\crefalias{section}{appendix}
\section{Query implementation and general tools}\label{app:model}

This appendix supplies the query tools outlined in \cref{sec:model},
followed by retry and confidence bounds, fractional plans with few split cells and
protocols for at most three agents. All query bounds include recording
new endpoints for all $N$ original agents (\cref{rem:cache-invariant}).

\subsection{Measure and probability conventions}\label{app:measure-conventions}

The formal valuation assumption behind \cref{sec:model} is that each
$v_i$ is a finite, nonnegative, nonatomic Borel measure on $[0,1]$, with
$v_i([a,b])>0$ whenever $a<b$. Thus the cumulative value
$F_i(x)=v_i([0,x])$ is continuous and strictly increasing. A feasible
interval cut is the unique point
$F_i^{-1}(F_i(a)+t)$. No density is assumed: hungriness requires positive
value on every interval, but does not require agents to agree on which
other measurable sets have value zero. Restrictions to the pieces used
by the protocol and differences of valuations remain finite measures,
the latter possibly signed.

For a fixed valuation profile, the random experiment consists of a
sequence of independent uniform real draws. Interval evaluations and
the unique cuts are measurable functions of their inputs. Arithmetic,
comparisons and the finite choices in the protocol preserve
measurability; ties in assignments, cell choices and linear algebra are
resolved by fixed orders. Consequently the records after each finite
step, the acceptance checks and the event of eventual termination are
measurable. Probabilities conditional on a history retain all preceding
observations, including records discarded after rejection. Fresh draws
are independent of this full history. These conventions justify the
conditional expectations below without imposing further regularity on
the valuations.

\subsection{Cuts and representation size}

Recall that a prefix of an ordered finite union is an initial segment in
its virtual order: whole components followed by part of the next,
skipping gaps. A prefix cut asks for the prefix on which an agent's value,
scaled by a known factor, equals a target; the next lemma implements it
with ordinary interval queries.

\Needspace{14\baselineskip}
\begin{lemma}[Cuts on pieces with recorded endpoint values]\label{lem:union-cut}
Let $S$ be an ordered finite union with recorded endpoint values.
\begin{enumerate}[label=(\alph*)]
\item Every agent's value of a union with endpoints in $\mathcal K$ is a
sum of differences of recorded values and needs no query.
\item For an original agent $a$, known $c>0$ and $y\in[0,c\,v_a(S)]$, a
prefix $S_x$ with $c\,v_a(S_x)=y$ can be found, and its endpoint values
recorded, with at most one interval $\Cut$ and $N$ $\Eval$ queries.
\item If $y$ is the cumulative $c\,v_a$-value at a component endpoint, the
procedure of (b) reuses the right endpoint of the preceding component
(or the first left endpoint when $y=0$ and $S$ is nonempty), and makes
no $\Cut$.
\item Otherwise it cuts from the left endpoint $l$ of the component
$[l,r]$ containing the target, at a residual target strictly between $0$
and $v_a([l,r])$. The unique reply lies in $(l,r)$.
\end{enumerate}
\end{lemma}

\begin{proof}
For recorded endpoints $x<x'$,
$v_j([x,x'])=v_j([0,x'])-v_j([0,x])$. Adding these differences over the
components proves (a).

For a cut, omit isolated endpoints, which have zero value. If $S$ is
empty, the only target is zero and the empty prefix suffices. Otherwise
write $[l_h,r_h]$ for its positive-length components and $\Sigma_h$
for the cumulative $c\,v_a$-value through component $h$, with
$\Sigma_0=0$. Hungriness makes these recorded values strictly increasing
in $h$. If $y=\Sigma_h$, return the first $h$ components, using the
empty prefix ending at $l_1$ for $h=0$. The endpoint is already recorded,
proving (c).
At a gap, the preceding component's right endpoint represents this
prefix; no cut across the gap is needed.

Otherwise there is a unique $h$ with $\Sigma_{h-1}<y<\Sigma_h$, and the
procedure asks $\Cut_a(l_h,(y-\Sigma_{h-1})/c)$. Subtracting the value of
the earlier components and dividing by the scaling factor leaves a
residual target strictly between zero and $a$'s value of the component.
Continuity and strict increase give a unique reply strictly inside
$[l_h,r_h]$, with the required prefix value, proving (d).
Recording the new endpoint $x$ with $\Eval_j(0,x)$ for $j\in[N]$
gives the bound in (b): at most one $\Cut$ and $N$ $\Eval$ queries.
\end{proof}

We store every piece using intervals of one common partition of the cake.
Taking unions, including unions of recursive outputs, therefore creates
no endpoints.

\begin{lemma}[Size of the working representation]\label{lem:additive-growth}
If the current working state contains $B_0$ recorded endpoints,
and the current attempt creates $C$ more, every stored piece is a union
of intervals from one partition of $[0,1]$ into at most $B_0+C+1$
intervals. Unions,
differences, complements, concatenations of component lists and
reassignments create no endpoints and cost no queries.
\end{lemma}

\begin{proof}
The recorded and new endpoints partition the cake into at most
$B_0+C+1$ intervals. Each listed operation only selects or reassigns
these intervals, up to finitely many points; it introduces no endpoints,
and \cref{lem:union-cut}(a) evaluates the result without queries.
\end{proof}

A logical cell is consecutive in the virtual order (\cref{def:cache}).
It may contain the end of one component and the start of the next: the
gap contributes no input cake or cumulative value
(\cref{rem:sampling-domain}). Stored endpoints support interval
operations, but only the boundaries prescribed by a trial divide its
logical sampling grid. Endpoints retained from earlier successful
attempts do not refine that grid.

\subsection{Retries and query caps}

The success probability must stay bounded below after every history.
This lets later requests depend on earlier outcomes.

\Needspace{10\baselineskip}
\begin{lemma}[Repeated trials]\label{lem:retry}
Suppose that trials are repeated until an exactly checked condition
holds, that each trial costs at most $C$ queries and succeeds with
probability at least $p>0$ conditional on every history at its start, and
that rejected trials allocate nothing. Then the expected numbers of trials
and queries are at most $1/p$ and $C/p$, respectively. At most $B$
requests for such successes, with adaptively chosen inputs, cost at most
$BC/p$ in expectation.
\end{lemma}

\begin{proof}
Let $T$ be the number of trials up to the first success. Since the
bound $p$ holds conditional on every history, it still holds after any
sequence of failures. Hence $\Pr[T>u]$, the
probability that the first $u$ trials all fail, is at most $(1-p)^u$, and
$\E T=\sum_{u\ge0}\Pr[T>u]\le1/p$. Multiplying by the per-trial query
bound gives $C/p$.

For requests chosen using earlier results, apply the bound conditional
on each starting history, then sum.
\end{proof}

A confidence bound needs more than an initial expected running time.
We show in \cref{sec:accounting} that the expected number of \emph{further}
attempts is bounded after every attempt, even if earlier
attempts have failed. The next lemma turns that property into a fixed
budget for the same continuing run. Successful work is retained.

\begin{lemma}[A confidence bound for a continuing run]\label{lem:continuation-cap}
Suppose a protocol terminates almost surely, returns only verified
outputs, and uses finitely many deterministic operations between
attempts and after its last attempt. At every reached boundary
before an attempt, conditional on the full history, suppose its
expected number of further attempts is at most a known $H$.
For any integer $B\ge2H$ and positive integer $r$, the probability
that more than $rB$ attempts are needed is at most $2^{-r}$.

Thus the same run may be stopped before attempt $rB+1$, reporting
failure if another attempt is needed. All deterministic processing and
final checks after attempt $rB$ are allowed.
\end{lemma}

\begin{proof}
Let $T$ be the total attempt count. After $jB$ attempts and their
deterministic processing, either the protocol has returned or its
expected number of further attempts is at most $H$. In the latter case,
conditional Markov gives
\[
 \Pr[T>(j+1)B\mid\text{history after }jB\text{ attempts}]
 \le H/B\le\tfrac12.
\]
Averaging over the histories on which $T>jB$ gives
$\Pr[T>(j+1)B]\le\Pr[T>jB]/2$. Induction proves the tail bound.
No independence between blocks is required.
\end{proof}

For a requested failure probability $\delta$, choose the least positive
integer $r$ with $2^{-r}\le\delta$. Repeated halving finds it in
$O(r)$ operations, and $r=O(1+\log(1/\delta))$. The budget is checked
before preparation for the next attempt, rather than in the
middle of a construction. Discarding an unsuccessful attempt's working
records does not erase its query charges or the oracle's history;
the conditional bounds allow that full history.

\subsection{Fractional assignments and small instances}

Reserve payments and simultaneous carving begin with a numerical plan:
each label receives the same prescribed fraction of every cell's value.
A fraction $x$ credits the label with $x$ times each agent's value of
that cell; it need not correspond to a physical subpiece with those values.
We adjust the fractions while preserving the required totals until few
cells are \emph{split}, with positive fractions for more than one label.
Discarding those cells leaves whole cells; the identity below accounts
for the lost value.

\Needspace{11\baselineskip}
\begin{lemma}[Few split cells]\label{lem:sparse-fractions}\label{lem:lp-vertex}
Suppose $H$ cells have nonnegative fractions $x_{hj}$ for labels
$j=0,\ldots,M$, with $\sum_jx_{hj}=1$ in each cell and $r$ additional
linear equations. Given a feasible plan and its coefficient table, a
deterministic algorithm finds a feasible plan with at most $r$ cells
having more than one positive fraction. It uses no valuation queries and
$O(H(M+1)(1+r^3))$ exact arithmetic operations, with polynomially many
list operations. Label $0$ may represent unassigned cake.
\end{lemma}

\begin{proof}
Suppose more than $r$ cells are split. Choose $r+1$ distinct split cells
and two positive fractions in each. In each chosen cell consider the
direction that increases one of these fractions by one and decreases the
other by one. This preserves its cell sum. The effects of these $r+1$
directions on the additional equations are vectors in $\mathbb R^r$, so
Gaussian elimination finds a nonzero combination whose effect is zero.
The combined direction $z$ is nonzero because the chosen cells are
distinct. It changes the fractions without changing any required total.
It has both positive and negative entries, so we can move until a
positive fraction first becomes zero:
\[
 \epsilon=\min_{z_{hj}<0}\frac{x_{hj}}{-z_{hj}},\qquad
 x\leftarrow x+\epsilon z.
\]
All changed fractions were positive. Consequently no zero fraction
becomes positive, no fraction becomes negative, and at least one positive
fraction disappears. Each cell still has fractions summing to one.
Repeat while more than $r$ cells are split.

There are initially at most $H(M+1)$ positive fractions, and at least
$H$ remain, so at most $HM$ moves are possible. Each solves an
$r$-by-$(r+1)$ homogeneous system and changes at most $2(r+1)$
coordinates. Lists of positive fractions and split cells supply these
coordinates. Forming the system, solving it and updating the fractions
costs $O(1+r^3)$ arithmetic operations per move. Initialisation costs
$O(H(M+1))$, proving the stated bound. The argument also covers $r=0$,
redundant equations and initially zero fractions. Fixed cell, coordinate
and pivot orders make the algorithm deterministic.
\end{proof}

An observer is an agent whose value must be preserved; a label names a
recipient or unused cake. The loss from discarding split cells is their
contribution to the numerical plan. Write $E_h$ for cell $h$ and
$\beta_{ij}$ for observer $i$'s target for label $j$. Suppose the imposed
equations include $\sum_h x_{hj}v_i(E_h)=\beta_{ij}$. Let $\mathcal D$
be the split cells and let $Y_j$ contain the cells assigned wholly to
$j$. Then
\begin{equation}\label{eq:rounding-residual}
 v_i(Y_j)=\beta_{ij}-\sum_{h\in\mathcal D}x_{hj}v_i(E_h),
 \qquad |\mathcal D|\le r,\qquad
 \sum_{j=1}^M x_{hj}\le1.
\end{equation}
The last inequality omits the unused label $0$. This identity also
applies to normalised valuations. With $k$ observers and $M$ recipient labels
there are $kM$ value equations, so at most $kM$ split cells remain.
Carving uses this directly; reserve payments have at most $k$ observers
and $k$ recipients, giving at most $k^2$ discarded cells.

Classical protocols handle the final calls, those with at most three
agents. Running them in the virtual order also handles cake with gaps.

\Needspace{8\baselineskip}
\begin{lemma}[Bounded protocols for at most three agents]\label{lem:base-cases}
For $1\le k\le3$ agents and a finite union $U$ with recorded endpoint
values, a deterministic procedure returns a complete envy-free allocation:
it gives $U$ to its sole agent, uses cut-and-choose for two agents, and
uses Selfridge--Conway for three \cite{RobertsonWebb1998}. These run in
the virtual order of $U$, using \cref{lem:union-cut} for each cut. The
cost is $O(N)$ queries.
\end{lemma}

\begin{proof}
The classical protocols use only evaluations of pieces and cuts of a
piece at a prescribed value from its start, both supplied in the virtual
order by \cref{lem:union-cut}; it remains to count queries. At most five
cuts are needed: two to divide $U$ into thirds, one trim, and two to
divide the trimmings (or one half-cut for two agents). By \cref{lem:union-cut}(b), each cut costs at most one cut query and
$N$ evaluation queries, while evaluating existing pieces costs no queries
by \cref{lem:union-cut}(a). Thus the total is $O(N)$. Each cut realises the exact value used in the classical
envy-freeness argument, and the virtual order preserves completeness.
\end{proof}
\section{Polynomial computation}\label{app:computation}

Polynomial computation requires both polynomial work on the recorded
values and a polynomial bound on their storage. We prove these in turn,
using the operation model in \cref{def:computation-model}.

\subsection{Representative cells and fractional plans}\label{sec:representative-computation}\label{sec:sparse-computation}

To choose the representative cells for repair, we use the two scans in
\cref{lem:representative-basis}. With $H$ available cells, they cost
$O(H)$ operations per pair.
Fewer than $k^2$ pairs in each of $L$ rounds, repeated for $k$ disjoint
batches, therefore cost $O(Lk^3H)$. As required by
\cref{sec:kernel-witnesses}, each selected cell's coarse parent is
removed immediately, before the next pair is considered. The scans use
input-normalised values, never reserve labels, and make no queries.

The fractional plans for payments and simultaneous carving are computed
by the elimination algorithm in \cref{lem:sparse-fractions}. Each has at most $k+1$ labels and
$k^2$ additional equations, giving $O(Hk^7)$ arithmetic operations.
The starting fractions are explicit, so neither application requires
valuation queries or a general linear-programming algorithm.

\subsection{Other operations and stored pieces}\label{sec:remaining-computation}

The remaining calculations are standard combinatorial algorithms or scans
of finite tables.

\smallskip\noindent
\emph{Random seed pieces.}
One uniform draw and at most $r$ comparisons with rational thresholds
choose uniformly among $r$ positions. This implements all pair thresholds
in $O(k^3L)$ operations, including the witness choices and inclusions.
Successively choosing among the remaining batch pieces gives an
independent uniform permutation in $O(k^2)$ operations and at most $k$
draws.

\smallskip\noindent
\emph{Assignments.}
A maximum-value assignment is a maximum-weight bipartite matching, and
the Hungarian method finds one with polynomially many exact arithmetic
operations and comparisons~\cite{Kuhn1955,Munkres1957}. Fix a deterministic
implementation, including its tie rules. The resulting optimum is the
fixed choice the protocol needs; no extra matching calls are required.

\smallskip\noindent
\emph{Credits.}
The credits of \cref{sec:kernel-credits} are longest-path values with
no positive-cost cycle. Add a source with zero-cost edges to every
agent and use the usual shortest-path relaxation algorithm with signs
reversed. It computes the credits or detects a positive-cost cycle,
in which case the trial rejects, using polynomially many operations.
The forced-recipient comparisons in \cref{lem:kernel-isolation}
belong to the analysis and are not performed by the protocol.

\smallskip\noindent
\emph{Completion.}
Completion stores at most $k^2$ snapshots per stage. Its graph searches
find strongly connected components, paths and the two final groups.
Each of at most $k(k-1)$ greedy steps scans one fresh snapshot and uses
a path to determine the reassignment; it need not enumerate permutations.
These operations are polynomial in the number of agents and stored pieces.

\smallskip\noindent
\emph{Stored pieces.}
Pieces are interval lists over the common endpoint partition of
\cref{lem:additive-growth}. Ordering endpoints, taking unions and
differences, locating a cut target, and checking coverage and all envy
inequalities are therefore polynomial in the stored table size and $N$.
The table retains all $N$ rows, including those of agents who have
stopped participating (\cref{rem:cache-invariant}).
The call schedule of \cref{prop:completion-stage} bounds the number of
retained pieces polynomially in $N$ and the number of successful
attempts.

\subsection{Working data and total cost}\label{sec:expected-computation}

\begin{proposition}[Polynomial expected computation]\label{prop:polynomial-work}
The continuous-randomness protocol on $[0,1]$ has polynomial expected
total operations in the model of \cref{def:computation-model}. Its
bounded-confidence version has an operation cap polynomial in $N$
multiplied by $1+\log(1/\delta)$. These bounds hold uniformly over
all hungry valuation profiles.
\end{proposition}
\begin{proof}
\emph{Size of the working data.}
Each attempt creates $O(s^{24})$ endpoints, including its sampled
tables, proposed pieces and any corrections. A snapshot adds $O(k)$
endpoints beyond at most two sampled tables; shrinking adds
$O(k^3s^{21})=O(s^{24})$ carving endpoints and one repair table. Completion
makes no other cuts except the $O(1)$ cuts in its final small-agent
case. Consequently, at every point there are at most
\[
 O\bigl((B_N+1)s^{24}\bigr)
\]
endpoints: $B_N$ accounts for successful attempts and the extra term for
the current attempt (\cref{lem:total-trials}). Each endpoint has $N$
recorded values. One
saved copy of the working state suffices for restoration; neither
temporary records nor saved copies accumulate after rejection.

\emph{Operations per attempt.}
Every stored piece is an interval list over this common endpoint
partition (\cref{lem:additive-growth}). Each successful attempt retains
only polynomially many named pieces, finite tables and graph objects;
the current attempt adds one further such collection. Thus the scans,
linear systems, matching, graph searches and exact checks above take
polynomially many operations per attempt. This includes copying and
restoring its state.

The random draws are polynomially bounded too. A repair uses one per
reserve label, two per mixture sample, one per pair threshold, at most
$k$ for the batch permutation, and one per positive correction deficit.
A snapshot has its outer samples and at most two circular starts,
followed by at most one repair.

The numerical schedules use rational arithmetic and fixed integer
powers. Shrinking reaches its target by repeated multiplication;
\cref{sec:accounting} computes the attempt allowance by fixed integer
arithmetic and the confidence budget by repeated halving. Consequently there are deterministic polynomials $W_0(N),W(N)$ such
that, apart from computing the confidence parameter, the first $Z$
attempts and their deterministic processing use at most
\[
 W_0(N)+W(N)Z
\]
operations. Here $W_0$ covers initialisation, final verification and the
finite completion decisions; $W$ includes processing after an attempt
and before the next one. Queries and cleanup after failed attempts
remain charged.
For a local call on an arbitrary finite union, its supplied interval
representation also contributes; the population-only bound starts on
$[0,1]$.

\emph{Expectation and confidence.}
By \cref{lem:total-trials}, $\E Z\le B_N$, giving polynomial expected
operations. The confidence version in \cref{sec:accounting} permits
$O(B_N(1+\log(1/\delta)))$ attempts and includes the final deterministic
processing. Since the working-data bound is independent of $\delta$,
the same polynomial cost per attempt proves the stated operation cap.
Computing the confidence parameter by repeated halving adds
$O(1+\log(1/\delta))$ operations, which this cap also covers.
\end{proof}
\section{Sampling with ordinary queries}\label{sec:sampling}

The trials of \cref{sec:kernel,sec:snapshot} compare agents' values on
whole grid cells, then cut inside them. We construct random points and
pieces of prescribed value using ordinary single-agent queries
(\cref{lem:sampling-average,lem:sampling-crop}), and control the errors
at later cuts relative to the differences already observed
(\cref{prop:sampling-endpoints}). Cuts may depend on
the samples, and valuations need not have densities.
Every queried endpoint is recorded for all $N$ original agents
(\cref{def:cache}).

The pieces to which the estimates are applied may be disconnected leftovers
of earlier divisions, and an error must not be charged at each gap. We
therefore treat such a union as a single ordered interval.

\begin{remark}[A fixed ordered domain]\label{rem:sampling-domain}
Fix an ordered finite union $D_0$ of positive length with recorded endpoint
values, and restrict all measures to it. Before sampling, omit degenerate components,
place the remaining intervals consecutively without gaps, and rescale
their total length to $[0,1]$. This fixes a length coordinate $\ell$
without changing valuations. A grid cell may cross several original
components. The finitely many joins carry no mass; cumulative value is
the same on either side of a join, so no error is charged there.
Cuts on this ordered union use \cref{lem:union-cut}.
\end{remark}

\subsection{Random points and pieces of prescribed value}

The sample points should follow the average of the agents' normalised
valuations, but a single query cuts according to one agent's valuation
only. We therefore randomise the agent as well: we choose an agent
uniformly and then cut at a uniform fraction of her value. For example,
a uniformly chosen value target falls in the first third of her value
with probability $1/3$, regardless of that prefix's physical length.

\Needspace{10\baselineskip}
\begin{lemma}[Sampling the average measure with one Cut]\label{lem:sampling-average}
Let $S$ be an ordered finite union of positive length with recorded
endpoint values, and $I$ a set of $k\ge1$ agents. Set $\nu_i=v_i|_S/v_i(S)$ and
$\mu=k^{-1}\sum_{i\in I}\nu_i$. Choose $i$ uniformly from $I$ and,
independently, $Y$ uniformly in $(0,1)$. Let $X$ be the point returned by
\cref{lem:union-cut} at target $Yv_i(S)$.
\begin{enumerate}[label=(\alph*)]
\item Given the history before $(i,Y)$, $X$ has distribution $\mu$.
\item Fresh repetitions give independent $\mu$-points.
\item Each draw costs at most one $\Cut$ and $N$ $\Eval$ queries.
\end{enumerate}
\end{lemma}

\begin{proof}
Fix the history and the chosen agent $i$. Place the components of $S$ consecutively as in
\cref{rem:sampling-domain}, and write $W_i(x)=\nu_i(S_x)$ for the value
of the prefix ending at $x$. This is continuous and strictly increasing
from $0$ to $1$, so the cut gives $X=W_i^{-1}(Y)$. Hence
\[
 \Pr[X\in S_x]=\Pr[Y\le W_i(x)]=W_i(x)=\nu_i(S_x).
\]
Component endpoints have probability zero. Prefix probabilities
determine the distribution, so averaging over the
chosen agent proves (a). For a fixed domain and agent set, each fresh draw still has
distribution $\mu$ after conditioning on the preceding draws. This
proves independence in (b). Part (c) is \cref{lem:union-cut}(b).
\end{proof}

Payments need random pieces of prescribed value. Controlling their boundary
distributions will let us bound the error caused by using the sampled
table. To obtain such a
piece we join the ordered components into a circle, skipping the gaps, and
take the requested amount from a uniformly random starting value. This
\emph{circular crop} favours neither end of the union. Its \emph{owner} is
the agent whose valuation supplies the cuts; she need not receive the crop.
Amounts are measured in the owner's value, not in length.

We also compare one agent's value with another's. Their difference
$\xi(E)=v_i(E)-v_j(E)$ may be positive on one piece and negative on
another; this is a \emph{signed measure}. Its \emph{total variation}
$|\xi|$ counts these differences without cancellation: it adds the
positive mass and the absolute value of the negative mass. The next
lemma allows any finite signed nonatomic measure, including normalised
differences and the cell estimates used later.

\begin{lemma}[Circular pieces and partitions]\label{lem:sampling-crop}
Fix, before the randomness, an ordered finite union $S$ with recorded
endpoint values, an owner measure $\omega=cv_a$ for an original agent $a$
and known $c>0$, and
amounts $y_1,\ldots,y_r\ge0$, where $r\ge2$ and
$\sum_h y_h=\omega(S)=t>0$. A fresh uniform starting value $\Upsilon\in[0,t)$
produces disjoint finite unions $C_1,\ldots,C_r$ covering $S$ up to finitely
many points, with the following properties.
\begin{enumerate}[label=(\alph*)]
\item For every starting value, $\omega(C_h)=y_h$ for every $h$.
\item Apart from component endpoints, there are at most $r$ boundaries,
  using at most $r$ $\Cut$ and $rN$ $\Eval$ queries. If some $y_h=t$,
  take $C_h=S$ and the other parts empty, without queries.
\item Every new boundary $Z$ has conditional distribution $\omega|_S/t$ given the
  history before $\Upsilon$ is drawn; boundaries may be dependent. In particular,
  if $\omega\le\gamma\mu$ for a probability measure $\mu$ and $t\ge t_0>0$,
  with deterministic $\gamma,t_0$, then
  $\Pr[Z\in E\mid\text{history}]\le(\gamma/t_0)\mu(E)$ for measurable $E$.
\item Except at component endpoints, each fixed $x\in S$ belongs to $C_h$
  with probability $y_h/t$. Thus, for any finite signed nonatomic measure
  $\xi$ fixed before $\Upsilon$ is drawn,
  \begin{equation}\label{eq:sampling-crop-expectation}
   \E[\xi(C_h)\mid\text{history}]=\frac{y_h}{t}\xi(S),
   \qquad
   \E[|\xi(C_h)|\mid\text{history}]
   \le\frac{y_h}{t}|\xi|(S).
  \end{equation}
  No absolute continuity between $\xi$ and $\omega$ is required.
\end{enumerate}
A \emph{crop of amount $y$} uses $r=2$, $y_1=y$ and $C=C_1$; for
$0<y<t$ it requires at most two new boundaries and two $\Cut$ queries.
\end{lemma}

\begin{proof}
\emph{The construction.} Give each zero amount an empty part. Relabel
the positive amounts in their original cyclic order for this proof,
and write $r$ for their number. A single positive amount is handled
as in (b). Otherwise join the two ends of $S$ into a circle and put
$s(x)=\omega(S_x)$ in its gap-free ordered coordinate, where $S_x$ is the prefix
of $S$ ending at $x$, and $\Upsilon_h=(\Upsilon+\sum_{j<h}y_j)\bmod t$.
Cutting $S$ at each target $\Upsilon_h$ gives a point $z_h$, and $C_h$ is
the forward path in $S$ from $z_h$ to $z_{h+1}$, continuing from the start
if needed, where $z_{r+1}=z_1$. The path skips gaps outside $S$. There
are at most $r$ cut points, each found with at most one $\Cut$ and $N$
$\Eval$ queries (\cref{lem:union-cut}(b)), and assigning each to one
adjacent part makes the parts disjoint. Restore the original labels
afterwards. Since omitting zero amounts only reduces the number of
boundaries, this gives (b).

\emph{Values of the parts.} The positive amounts give distinct target
values around the circle. Monotonicity puts their cut points in the same
circular order. In value coordinates, $C_h$ traces the arc of length $y_h$ from
$\Upsilon_h$ to $\Upsilon_{h+1}$, and it passes through the start of $S$
exactly when this arc wraps around the circle; in either case its value
changes add up to $y_h$. This proves (a).

\emph{Distributions of the boundaries.} Given the history before
$\Upsilon$, each $\Upsilon_h$ is uniform on $[0,t)$. Applying the
inverse-value argument of \cref{lem:sampling-average} to the strictly
increasing function $s$ gives $\Pr[z_h\in S_x]=s(x)/t$.
Thus each boundary has distribution
$\omega|_S/t$, and (c) follows from
$\omega(E\cap S)/t\le\gamma\mu(E)/t_0$.

\emph{Inclusion probabilities.} Fix $x$ away from component endpoints.
Except on the null event that
$s(x)$ equals a boundary level, $x\in C_h$ exactly when $s(x)$ lies in the
uniformly rotated arc of length $y_h$. The probability of
this event is $y_h/t$. The finitely many component endpoints have zero
$\xi$-mass. Integrating against each of the nonnegative
positive and negative parts of $\xi$ gives the expectation identity by
Tonelli's theorem. Applying the same identity to $|\xi|$ and using
$|\xi(C_h)|\le|\xi|(C_h)$ gives the inequality.
\end{proof}

The amounts must be fixed before the random start: choosing
$y_1=t-\Upsilon$ would force a boundary at the start. Part (c) also
conditions before the shared draw: revealing one boundary can determine
the starting value, and hence the other targets.

\subsection{Sampling error for later pieces}

The table records differences between agents on whole cells. Later cuts
can expose differences inside those cells. We need to bound this extra
error relative to the differences already recorded: a fixed error
allowance could erase a small preference.

We first give a simple estimate for part of a cell. If a piece occupies
one tenth of the cell's length, give it one tenth of the recorded
difference. Equivalently, join cumulative values by a straight line in
the length coordinate between recorded boundaries. This linear
interpolation is an analytical estimate, not an
assumption that actual valuations are uniform inside cells. The
proposition below controls its error even when later pieces are chosen
after inspecting the table.

\begin{definition}[Finite cell estimates]\label{def:sampling-model}
Fix a nonatomic probability measure $\mu$ on the ordered domain $[0,1]$,
a finite set of permanent boundaries $\mathcal F\ni0,1$, and an integer
$m\ge0$. Adjoin $m$ independent $\mu$-points to obtain the sampled grid $P$.
For a finite signed nonatomic measure $\sigma$ fixed before sampling,
write $\delta_{Q'}=\sigma(Q')$ for its recorded value on a grid cell and
$D_\sigma=\sum_{Q'}|\delta_{Q'}|$ for the total absolute recorded value. The
\emph{cell estimate} is
\begin{equation}\label{eq:sampling-model}
 \widehat\sigma(E)=
 \sum_{Q'}\delta_{Q'}\frac{\ell(E\cap Q')}{\ell(Q')}.
\end{equation}
Here the sum runs over the cells between consecutive distinct grid
points, each of positive length. A typical comparison is
$\sigma(E)=c_i\nu_i(E\cap A)-c_j\nu_j(E\cap A)$, with both valuations,
coefficients and the region $A$ fixed before sampling. A single
valuation is also allowed.
\end{definition}

The number $D_\sigma$ measures the differences visible on the grid;
opposite differences inside a cell may cancel. The estimate adds over
disjoint pieces, agrees with recorded values on whole cells, and is linear
in $\sigma$. In particular,
$\widehat{\nu_i-\nu_j}=\widehat\nu_i-\widehat\nu_j$. Its total variation is
\begin{equation}\label{eq:sampling-model-absolute}
 |\widehat\sigma|(E)
 =\sum_{Q'}|\delta_{Q'}|\frac{\ell(E\cap Q')}{\ell(Q')},
 \qquad |\widehat\sigma(E)|\le|\widehat\sigma|(E)\le D_\sigma.
\end{equation}
Thus $|\widehat\sigma|([0,1])=D_\sigma$. If $\sigma([0,1])=0$, its positive
cell values sum to $D_\sigma/2$, as do the absolute values of the negative
ones. In that case $|\widehat\sigma(E)|\le D_\sigma/2$, with equality for
the union of positive-value cells and its complement.

A circular piece also has a simple estimated value. Fix the grid,
cutting region $S$, owner and requested value $y$ before a fresh start,
and let $t>0$ be the owner's value of $S$. Applying
\cref{lem:sampling-crop}(d) to the signed measure $\widehat\sigma$ gives
\begin{equation}\label{eq:sampling-model-crop}
\begin{aligned}
 \E[\widehat\sigma(C)\mid\text{history}]&=\frac yt\widehat\sigma(S),\\
 \E[|\widehat\sigma(C)|\mid\text{history}]
 &\le\frac yt|\widehat\sigma|(S)\le\frac ytD_\sigma.
\end{aligned}
\end{equation}
The estimate and the owner's valuation need not agree on which
measurable sets have value zero; the inclusion identity applies to both.
If earlier pieces have been removed,
restrict the \emph{original} estimate to the remaining region $S$;
do not interpolate again on a different grid.

To control cuts chosen after inspecting the table, we bound how heavily
new endpoints can concentrate in any region. Circular cuts generate their
endpoints together, so we state the condition for groups sharing a fresh
random choice.

\begin{definition}[Groups of later boundaries]\label{def:sampling-slots}
Fix an integer $H\ge0$ and a constant $\Lambda>0$ before sampling.
After the sample grid $P$ is drawn, generate a list $\mathcal Z$ of at
most $H$ new endpoints in groups. Each group's size is fixed before its
fresh random choice. Given the full history $\mathcal H$ before that
choice, including $P$, every endpoint $Z$ in the group must satisfy
\begin{equation}\label{eq:sampling-boundary-count}
 \Pr[Z\in E\mid\mathcal H]\le\Lambda\mu(E).
\end{equation}
This is an almost-sure inequality between conditional measures, for
all measurable regions $E$. In particular it applies to regions and
nonnegative test functions determined by $P$. Group choices and stopping
may depend on earlier outcomes; endpoints within a group need not be
independent. For a circular crop, condition before its shared random
start, not after revealing another boundary of that crop.

Permanent and sampled grid boundaries are not included: their estimation
error is already zero. A generated boundary may be reused without being
counted again; counting repeated points more than once is also harmless.
Include every boundary generated before stopping, including those of
the piece that causes rejection. An inherited boundary of a later piece
must already be in $P$ or in this list. Recording these endpoints does
not change the original cell estimate.
\end{definition}

\Needspace{12\baselineskip}
Whole cells have their exact recorded values, so a later piece incurs
error only at its boundaries. The next proposition bounds the sum of
these errors over all generated boundaries. This controls every later
piece at once, without an advance list of the pieces the protocol may
choose.

\begin{proposition}[Sampling error for adaptive pieces]\label{prop:sampling-endpoints}\label{cor:sampling-unions}
Fix $\mu,\mathcal F,m$ as in \cref{def:sampling-model}, a family of $J$
finite signed nonatomic measures, an integer $H\ge0$ and constants
$\Lambda,b>0$, all before sampling. Suppose the later endpoints satisfy
\cref{def:sampling-slots} with $H,\Lambda$. There is a joint failure event
$\mathcal B$ with
\begin{equation}\label{eq:sampling-endpoint-failure}
 \Pr[\mathcal B]\le
 \frac{2JH\Lambda(1+b)}{b(m+1)}.
\end{equation}
Outside this event, for every measure $\sigma$ in the family and every
finite union $E$ whose boundaries lie in $P$ or among the generated endpoints,
\begin{equation}\label{eq:sampling-union-error}
 |\sigma(E)-\widehat\sigma(E)|\le bD_\sigma.
\end{equation}
The union may be chosen after observing all endpoints and values. When
$D_\sigma=0$, the same event gives exactly zero error. One difference for
each unordered pair of agents covers both orientations.
\end{proposition}

\begin{proof}
\emph{The error at one boundary.} Fix one measure $\sigma$. Write
$g(x)=\sigma([0,x])$, and let $\widehat g$ join its recorded values by
straight lines. Set $D(P)=D_\sigma$ and
\[
 e(P,x)=|g(x)-\widehat g(x)|,\qquad
 r(P,x)=\frac{e(P,x)}{D(P)+e(P,x)},\qquad 0/0=0.
\]
Dividing by $D(P)+e(P,x)$ keeps the ratio bounded and still detects
positive error when the recorded variation is zero.

The original grid contains the permanent boundaries and $X_1,\ldots,X_m$.
Add one independent sample $X_0$, then test each of the $m+1$ sample points
against the grid formed by deleting it.
Let $V$ be the sum of the absolute cell values in this enlarged grid.
Almost surely the samples are distinct and avoid permanent boundaries.
For each $X_j$ with neighbours $L_j,R_j$, put
\[
 A_j=|g(X_j)-g(L_j)|+|g(R_j)-g(X_j)|,
 \qquad B_j=|g(R_j)-g(L_j)|.
\]
Deleting $X_j$ leaves observed variation $D'=V-A_j+B_j\ge0$.
The new interpolation error $e'$ at $X_j$ is at most $A_j$, because the
interpolated value lies between $g(L_j)$ and $g(R_j)$. Hence
\[
 \frac{e'}{D'+e'}\le
 \frac{A_j}{D'+A_j}=\frac{A_j}{V+B_j}.
\]
These inequalities also hold when a denominator is zero, using $0/0=0$.
Each cell of the enlarged grid contributes to $A_j$ only at its two ends,
so $\sum_j A_j\le2V$. If $V>0$, the sum of the deleted-point ratios is
therefore at most two; if $V=0$, all the errors are zero.
Every labelled sample has the same expected deleted-point ratio.
Deleting $X_0$ leaves the original grid $P$ and an independent
$\mu$-point $X_0$. Averaging the sum bound therefore gives
\[
 \E\!\int r(P,x)\,d\mu(x)=\E r(P,X_0)\le\frac{2}{m+1}.
\]
This is the insertion--deletion sampling principle
\cite{GaertnerWelzl2001}.

\emph{All later endpoints.} The function $r(P,\cdot)$ is nonnegative and
known before each fresh group choice. By
\eqref{eq:sampling-boundary-count}, a group of size $h$ contributes
conditional expected sum at most $h\Lambda\int r(P,x)\,d\mu(x)$.
Sum over the reached groups and use their total size at most $H$.
Conditional expectation, first given $P$ and then averaged, gives
\[
 \E\sum_{Z\in\mathcal Z}r(P,Z)
 \le H\Lambda\E\!\int r(P,x)\,d\mu(x)
 \le\frac{2H\Lambda}{m+1}.
\]
No independence among later endpoints is needed. Put
$E_* =\sum_{Z\in\mathcal Z} e(P,Z)$. Each denominator
$D(P)+e(P,Z)$ is at most $D(P)+E_*$, so, with the same
$0/0=0$ convention,
\[
 \sum_Z r(P,Z)\ge\frac{E_*}{D(P)+E_*}.
\]
Thus $E_*>bD(P)$ forces this sum to exceed $b/(1+b)$, also when $D(P)=0$.
Markov's inequality bounds this event by
$2H\Lambda(1+b)/(b(m+1))$. The union of these events over the $J$ measures
is $\mathcal B$, proving \eqref{eq:sampling-endpoint-failure}.

\emph{From boundaries to pieces.} The functions $g$ and $\widehat g$
agree on $P$. Merge adjacent intervals of $E$ and list its distinct
boundaries as $z_1,\ldots,z_{2a}$. Each interval's measure is the
difference of its cumulative values at its two ends, so
\[
 |\sigma(E)-\widehat\sigma(E)|
 \le\sum_{u=1}^{2a}|(g-\widehat g)(z_u)|
 \le E_*\le bD_\sigma
\]
outside $\mathcal B$. This holds for every union with allowed boundaries,
including choices made after all values are known. Nonatomicity makes
endpoint membership immaterial. Replacing $\sigma$ by $-\sigma$ changes
neither $D(P)$ nor $e(P,x)$, so the same event controls both orientations.
\end{proof}

\begin{remark}[Joint, not conditional]\label{rem:sampling-quantifiers}
The probability estimate averages over the initial sample and the later
cuts together. Some sampled tables can therefore give poor estimates;
the bound need not hold conditional on a particular table or a successful
output.
For example, take densities $3/2$ and $1/2$ on
$(1/4,1/2)$, exchanged on $(1/2,3/4)$, and both $1$ elsewhere. Their
average is uniform. With $\mathcal F=\{0,1\}$, sample points missing
$(1/4,3/4)$ observe zero variation of the difference, although a fresh
uniform point has positive endpoint error with probability $1/2$. Thus,
when success is judged by the cell estimate, what we obtain is
$\Pr[\text{success and invalid output}]\le\Pr[\mathcal B]$, not the same
bound on the conditional probability of failure given success.
Later owners, regions and targets may depend on the samples, provided
\cref{def:sampling-slots} still holds.
\end{remark}

\begin{remark}[Independent auxiliary choices]\label{rem:sampling-auxiliary}
The comparison measures may also be fixed by auxiliary data independent
of the samples, conditional on the starting history. Condition on these
data, apply the proposition, then average. This is valid when the family
has at most $J$ members and the same $H,\Lambda,b$ work with the auxiliary
data included in the history in \cref{def:sampling-slots}. The table may
select which already covered comparison to use; it may not define a new
comparison measure or reserve.
\end{remark}

The applications use the following bounds, verified in
\cref{sec:kernel,sec:snapshot}. Snapshot uses its single-candidate
construction.
\begin{center}
\begin{tabular}{lccc}
Application & $J$ & $H$ & $\Lambda$ \\
\hline
Repair corrections & $\le k^2/2$ & $\le2k$ & $4k/q$ \\
Snapshot starting pieces & $k-1$ & $\le2k$ & $2k^2$
\end{tabular}
\end{center}
For repair, fix the independent coarse reserve labels first. Labels
passing the mass check fix the reserve-normalised comparison measures
before sampling; apply the proposition and average over the labels.
Labels failing that check generate no correction endpoints. For Snapshot,
the cutter is fixed before sampling, so all $k-1$ comparisons with her
are fixed even though the table selects which one to use.

\begin{remark}[Query cost]\label{rem:sampling-cost}
In the applications (\cref{sec:kernel,sec:snapshot}), the sampling distribution and the valuation differences are
known combinations and restrictions of the original valuations. On a domain
with recorded endpoint values, $m$ sample points and at most $H$ later
endpoints cost at most $m+H$ $\Cut$ and $N(m+H)$ $\Eval$ queries, besides
other operations (\cref{lem:sampling-average}(c), \cref{lem:union-cut}(b)).
Actual values and cell estimates then use only recorded values and
length coordinates, with no further queries.
\end{remark}
\section{Numerical choices}\label{sec:app-kernel}

All numerical choices come from one scale, fixed by the original
number of agents. They serve three purposes. There must be enough
distinct cells to separate competing assignments; the reserve must
rarely contain a selected cell; and each cell must be small enough
that the resulting credits and payment errors fit within their
allowances. This appendix checks those requirements together with the
advantage needed for completion.

\Needspace{14\baselineskip}
\subsection{Repair parameters}

\begin{definition}[Parameter choices for repair]\label{def:kernel-parameters}
Let $N$ be the original number of agents, and fix
$s=2^{20}\max\{4,N\}$ for the whole execution. Every call, with any
$2\le k\le N$ participants, uses
\[
\begin{array}{c@{\qquad}l}
 \lambda=s^4/8 & \text{margin multiplier for payment errors},\\
 L=s^8 & \text{witness rounds for each pair in each batch},\\
 q=1/(s^3L)=s^{-11} & \text{probability of reserving a coarse cell},\\
 \eta=\eta_E=q/(s^4L)=s^{-23} & \text{maximum cell value relative to the input},\\
 m=s/\eta=s^{24} & \text{number of sample points}.
\end{array}
\]
The repair tolerance follows from the reserve scale:
\begin{equation}\label{eq:kernel-schedule}
 \tau=\frac{q}{4s^2}=\frac1{4s^{13}}.
\end{equation}
It is measured relative to the available cake. The mass check makes the
reserve worth at least a $q/2$ fraction of that cake to every
agent. After normalising the reserve to value one, another endowment can
therefore exceed her own by at most $2\tau/q=1/(2s^2)$. No bound is
needed in the reverse direction.

The following bounds are fixed in advance as well:
\begin{equation}\label{eq:kernel-derived}
 N_*=1024qm=1024s^{13},\qquad
 \eta_R=\frac{2\eta}{q}=2s^{-12},\qquad
 b=\frac{\lambda}{4N_*}=\frac1{32768s^9}.
\end{equation}
Here $N_*$ bounds the number of refined reserve cells, and $\eta_R$
bounds one cell's value in reserve units. The relative allowance $b$
controls errors at new cut endpoints. Since the total recorded
disagreement is at most $N_*A_{ij}$, an error of at most $b$ times that
total is at most $\lambda A_{ij}/4$.

The coarse grid has $1/\eta\le M_0\le2k/\eta\le m$ cells. Neither the
sample count nor any numerical choice depends on the realised grid or
on later reductions in the number of participants.
\end{definition}

\subsection{Repair and completion allowances}

The deterministic inequalities below ensure that credits are
affordable and that each successive correction leaves enough reserve
for the next one. They also bound the total cake added during repair:
this is the property used in the completion argument. The probability
bounds then control how often a trial has to be repeated.

\begin{lemma}[Numerical requirements]\label{lem:app-kernel-closure}
For the choices in \cref{def:kernel-parameters}, $0<q,b<1$ and
\[
\begin{array}{l@{\qquad}l}
 k^3L\eta<1/2 & \text{representatives remain available},\\
 k^2L\eta_R\le1/(4s^2) & \text{each random addition is small},\\
 2\lambda\eta_R\le1/(4s^2) & \text{each extra credit demand is small},\\
 k^2\le\lambda/2 & \text{whole-cell error fits its allowance},\\
 k\lambda\eta_R<1/16
     & \text{completed payments leave enough reserve}.
\end{array}
\]
If $0\le t_i\le k/s^2$, then
$t_i<1/(16k)$ and $\sum_i t_i<1/16$.
Moreover,
\[
 k^3L\eta+\frac{3q}{2}<\frac14,
 \qquad \tau<\frac34.
\]
The first inequality bounds repair's total additions by a quarter of
the input buffer in each participant's valuation. The second ensures
that the surviving completion advantage, $(7/4-\tau)$ times the
reference residue, exceeds the whole reference residue.

The five contributions to a repair trial's failure probability satisfy
\[
\begin{array}{rcl@{\qquad}l}
 4k\eta/q &<&1/512 & \text{reserve values},\\
 2qm/N_* &=&1/512 & \text{reserve-cell count},\\
 qk^3L &<&1/512 & \text{a reserved witness parent},\\
 \displaystyle\frac{k^3(12k\lambda+1)}{L+1}
     &<&1/512 & \text{assignment separation},\\
 \displaystyle\frac{8k^4(1+b)}{qb(m+1)}+\frac{2k^3\eta_R}{b}
     &<&1/512 & \text{the two correction errors}.
\end{array}
\]
Their sum is less than $5/512<1/64$. In particular, repair succeeds
with probability greater than $63/64$ after every starting history.
\end{lemma}
\begin{proof}
Throughout, $k/s\le2^{-20}$ and $s\ge2^{22}$. The deterministic
requirements follow from the substitutions
\[
\begin{array}{c@{\quad}c@{\quad}c}
 k^3L\eta=k^3/s^{15},&
 k^2L\eta_R=2k^2/s^4,&
 2\lambda\eta_R=1/(2s^8),\\
 \lambda/2=s^4/16,&
 k\lambda\eta_R=k/(4s^8),&
 \sum_i t_i\le k^2/s^2.
\end{array}
\]
Indeed, $8k^2\le s^2$ gives the random-addition bound; the same
scale gives $k^2\le\lambda/2$ and
$16kt_i\le16k^2/s^2<1$. The total-addition allowance is
$k^3/s^{15}+3/(2s^{11})<1/4$, and $\tau<3/4$.
The remaining displayed bounds are immediate from these substitutions.

For the failure probabilities, substitution gives the following
bounds. The last two rows together are the correction contribution.
\[
\begin{array}{l@{\qquad}l}
 \text{reserve values}&4k/s^{12}<1/512,\\
 \text{reserve-cell count}&1/512,\\
 \text{reserved witness parent}&(k/s)^3<1/512,\\
 \text{assignment separation}&(3/2)(k/s)^4+k^3/s^8<1/512,\\
 \text{boundary interpolation}&2^{19}(k/s)^4,\\
 \text{estimated correction value}&2^{17}(k/s)^3.
\end{array}
\]
Here the interpolation row uses $1+b\le2$ and $m+1\ge s^{24}$.
The last two rows sum to at most $2^{-61}+2^{-43}<1/512$.
Thus the five contributions total less than $5/512<1/64$.
\end{proof}
\subsection{The advantage relative to the remaining cake}\label{sec:app-snapshot}
\label{sec:app-snapshot-parameters}

Both positive-disagreement branches need an advantage that is an
inverse-polynomial fraction of the remaining cake. This subsection
checks that the same margin works for both and that the sampling
and query allowances cover their constructions.

\begin{definition}[Parameter choices for the snapshot procedure]\label{def:snapshot-parameters}
For $k\ge4$, use the standard choices of \cref{def:kernel-parameters}:
the global base $s=2^{20}\max\{4,N\}$, fixed throughout the execution,
mesh $\eta=\eta_E=s^{-23}$, tolerance $\tau=1/(4s^{13})$ and sample
count $m=s^{24}$. Also set
\[
 K=8/\tau=32s^{13},\qquad \varepsilon=s^{-2}.
\]
For $0<D<\tau/16$, the buffer has cutter-value $KD<1/2$.
The parameter $\varepsilon$ is both the relative sampling accuracy and
its failure allowance. The large-disagreement guarantee determines the
common margin below; the next lemma checks that the small-disagreement
case also meets it. For every participating count, use
\begin{equation}\label{eq:theta-snapshot}
 \theta=\frac{\tau}{768m}=\frac1{3072s^{37}}.
\end{equation}
These choices precede the observed disagreement $D$ and do not change
when the participating group becomes smaller. Only the actual buffer
value depends on $D$.
\end{definition}

The following bounds ensure that the starting pieces pass the repair input
condition, the measured preference survives the sampling error, and either
positive-disagreement branch supplies the stated margin.
\begin{lemma}[Snapshot scales]\label{lem:snapshot-scales}
The choices above satisfy
\[
 K\ge6,\qquad K-2\ge4/\tau,\qquad
 2\varepsilon\le\frac1{16k},\qquad
 \frac1{384k(K+2)}\ge\theta.
\]
If $D\ge\tau/16$ and $M\le2m$, then $D/(24M)\ge\theta$.
\end{lemma}
\begin{proof}
The buffer inequalities follow from $\tau<1$. The accuracy inequality
follows from $s^2\ge32k$. For the small-disagreement margin,
\[
 \tau k(K+2)=k(8+2\tau)\le10k\le2m,
 \qquad \frac1{384k(K+2)}\ge\frac{\tau}{768m}=\theta.
\]
For large disagreement,
\[
 \frac{D}{24M}\ge\frac{\tau}{768m}
 =\theta.
\]
All comparisons use $s\ge2^{22}$ and $k/s\le2^{-20}$.
\end{proof}

\paragraph{Enough sample points.}\label{sec:app-snapshot-menu}
The cutter $c$ is fixed before sampling. The maximising agent
$i^*$ is chosen afterwards, but the sampling argument only needs
comparisons of each agent with $c$. It therefore suffices to apply the
sampling estimate simultaneously to the $k-1$ fixed signed measures
\[
 \sigma_j=(v_j-v_c)|_U\qquad(j\ne c).
\]
The small-disagreement construction contributes at most $k+2\le2k$
new endpoints. Before each
fresh circular start, every endpoint it generates has conditional
distribution at most $2k^2\mu$ (\cref{lem:snapshot-menu}). Thus the
construction satisfies \cref{def:sampling-slots} with
\[
 H\le2k,\qquad \Lambda=2k^2.
\]
At zero disagreement, the single circular partition adds at most $k$
endpoints, each with conditional distribution $v_c\le k\mu$, so the same
bound applies. Every endpoint is counted, including those generated
before rejection. The original cell estimate stays fixed when these
endpoints are recorded. Inherited boundaries are either original table
boundaries or endpoints already on this list.

\Cref{prop:sampling-endpoints} now gives one joint exception
$\mathcal B_{\rm geo}$ for the snapshot table and later cuts, with
\begin{equation}\label{eq:snapshot-geo-failure}
 \Pr[\mathcal B_{\rm geo}]
 \le\frac{8k^4(1+\varepsilon)}{\varepsilon(m+1)}
 \le\frac{16k^4}{s^{22}}<\varepsilon.
\end{equation}
The last inequality follows from $s^{20}>16k^4$. By
\cref{cor:sampling-unions}, outside this exception every generated piece
satisfies the relative error bound. In particular, if $D_j=0$, agent
$j$ and the cutter agree exactly on each constructed piece; if $D=0$,
this holds for every agent. The bound averages over the table and later
cuts, as required in \cref{prop:snapshot-cost}.

\paragraph{Sizes and query count.}\label{sec:app-snapshot-cost}
\begin{remark}[Explicit sizes]\label{rem:snapshot-sizes}
One snapshot attempt uses $O(Ns^{24})$ queries. A returned snapshot
costs $O(Nks^{24})$ queries in expectation for $k$ input agents. Moreover
$\log(1/\theta)=O(\log(2N))$.
\end{remark}
\begin{proof}
Since $M_0\le2k/\eta_E\le m$, the table construction creates at most
$M_0+m\le2m$ grid endpoints and sample points. The small-disagreement
construction adds at most $k+2$ endpoints from two circular starts;
the zero case uses one start, and the large-disagreement input
uses recorded cells. Each new endpoint requires at most one cut and
$N$ evaluations. Thus constructing the repair input costs $O(N(m+k))=O(Ns^{24})$
queries. Its at most one fresh repair has the same allowance.
\Cref{prop:snapshot-cost} gives at most $64k$ attempts in
expectation, proving the returned-snapshot bound. Finally,
$\log(1/\theta)=\log3072+37\log s=O(\log(2N))$.
\end{proof}
\section*{AI methodology}
\pdfbookmark[1]{AI methodology}{ai-methodology}
This work developed through sustained exchanges between the authors and AI
systems. The authors supplied intuitions, questions and priorities, and directed
choices about the mathematical goals and how the arguments should be explained.
The AI systems developed and tested concrete realisations, proposed alternatives,
and substantially shaped and simplified the proofs. The process involved repeated
criticism, unsuccessful approaches, corrections to proposed arguments, and
revisions of both the mathematics and its presentation.

Anthropic's Claude Fable 5.1 and Claude Opus 5.5, used through Claude Code, and
OpenAI's GPT-6 Astra and GPT-6.1 Sol, used through Codex, contributed to proof
development and criticism,
manuscript drafting and revision, and checks of mathematical arguments and source
references. Their assistance substantially shaped the explanations in
\cref{sec:intro,sec:model,sec:overview}, the partial-allocation, margin and
completion constructions and their analysis in
\cref{sec:kernel,sec:snapshot,sec:completion}, the query accounting, and the
supporting proofs in the appendices.
GPT-6 Astra in Pro mode, used through ChatGPT, supplied feedback and proposed the polynomial computation
methods and standard-grid contrast argument. Codex assisted in
reconstructing, correcting and simplifying these proposals. Google Gemini feedback on the high-level explanation
and presentation informed the organisation of the exposition.
Codex also assisted in developing and checking the pairwise representative
selection, direct buffer construction, simpler compensation and numerical
choices. The models used through Codex also supported
literature and attribution review, mathematical reconstruction, proof
simplification and writing.

\end{document}